# Multi-pulse accumulation of gas molecular coherence enables gigahertz ultrafast frequency conversion

YAZHOU WANG,[1,*] MARCELLO MENEGHETTI,[1] IAN DAVIDSON,[2] TIMOTHY BATE,[3] J. E. ANTONIO-LOPEZ,[3] QIANG FU,[2] JAROSLAW RZEGOCKI,[2] GREGORY T. JASION,[2] RODRIGO AMEZCUA-CORREA,[3] FRANCESCO POLETTI,[2] AND CHRISTOS MARKOS[1,4,*]

[1]*DTU Electro, Technical University of Denmark, 2800 Kgs. Lyngby, Denmark*
[2]*Optoelectronics Research Centre, University of Southampton, Southampton, UK*
[3]*CREOL, The College of Optics and Photonics, University of Central Florida, Orlando, Florida 32816, USA*
[4]*NORBLIS ApS, Virum, Denmark*

**yazwang@dtu.dk; chmar@dtu.dk*

**Abstract:** Frequency conversion of ultrafast lasers is fundamentally constrained by the trade-off between pulse energy and repetition rate. Here, we demonstrate a frequency conversion mechanism in which molecular coherence accumulates across ultrafast pulse trains in gas-filled hollow-core fibers. Unlike conventional nonlinear interactions initiated by individual high-energy pulses, this mechanism relies on the collective buildup of coherent molecular oscillations driven by successive pulses. Using this approach, we achieve Raman frequency conversion at repetition rates up to 3 GHz with nanojoule pulse energies. The results demonstrate that accumulated molecular coherence can govern nonlinear frequency conversion, revealing a domain of light-matter interaction whose physical limits remain largely unexplored.

## 1. Introduction

Optical frequency conversion lies at the heart of ultrafast science but is fundamentally constrained by a trade-off between pulse energy and repetition rate. This limitation presents a long-standing challenge: when gigahertz-repetition-rate ultrafast lasers are frequency-converted across the ultraviolet to infrared spectral range, pulse energies are typically restricted to a few hundred picojoules or less [1–7], with only limited exceptions in parts of the near-infrared and visible regions [8–10]. Access to such sources at application-relevant wavelengths is nevertheless critical for frequency-comb spectroscopy [11], nonlinear microscopy [12,13], laser surgery [14,15], precision micromachining [14,16], free-space optical communication [17] and generation of high-brightness electron sources [18].

In conventional nonlinear frequency conversion schemes, this trade-off arises from the operational limits of the conversion medium. In nonlinear crystals, widely used for frequency conversion, gigahertz operation typically restricts pulse energies to <1 pJ in the ultraviolet and <200 pJ in the mid-infrared [3,5]. Such low pulse energies result from both limited pump energies and thus low conversion efficiencies. Increasing the pump energy however raises the average power and induces localized thermal effects that ultimately exceed the damage thresholds of existing materials [19]. Gases as alternative nonlinear media have attracted increasing interest. However, even with cavity-enhanced gas-based harmonic frequency conversion, pulse energies remain below tens of picojoules with repetition rates less than ~250 MHz because further scaling of repetition rate leads to plasma accumulation in the interaction region [20,21].

Gas-filled hollow-core fibers (HCFs) provide an alternative route by enhancing light–gas interaction over extended lengths. In particular, silica anti-resonant hollow-core fibers (ARHCFs) enable low-loss propagation across a broad spectral window from the ultraviolet to the mid-infrared while supporting average powers up to ~2 kW [22–30]. Despite these advantages, exploiting nonlinear effects in gas-filled ARHCFs typically requires pump pulse energies exceeding tens of microjoules [31–34]. As a result, repetition rates remain limited to a few megahertz, since scaling repetition rate to the gigahertz regime would require average powers approaching or exceeding the damage threshold of ARHCFs. One promising strategy to reduce the pump-energy requirement is the use of molecular coherence, in which optically prepared molecular coherence can enable frequency conversion of a weak probe optical signal, down to the single-photon level [35–40]. However, even in gas-filled hollow-core fibers, such approaches rely on high-energy pump pulses, typically at the microjoule level, to prepare the required molecular coherence [36,40].

Due to the fundamental limitations above, overcoming the trade-off between pulse energy and repetition rate requires a different optical frequency conversion mechanism. Here we demonstrate that multi-pulse accumulation of coherent molecular oscillations (MACMO) in gas-filled ARHCFs enables frequency conversion of ultrafast lasers with both nanojoule pulse energies and gigahertz repetition rates. Unlike conventional pump–probe approaches [35–40], the MACMO mechanism does not require a dedicated pulse for preparation of molecular coherence or a separate probe field. Instead, it relies on the cumulative build-up of molecular coherence waves by successive pump pulses, such that each pulse simultaneously contributes to coherence preparation and experiences the resulting modulation. This self-reinforcing process mitigates the need for high pump energy pulses and enables frequency conversion at gigahertz repetition rates. The MACMO mechanism is first demonstrated through Raman frequency conversion of a gigahertz repetition rate continuous pulse train, and its underlying dynamics are subsequently revealed through burst-mode pulses.

## 2. Concept

Figure 1(a) illustrates the concept. Compared to conventional stimulated Raman scattering (SRS), which relies on strong excitation by individual high-energy pulses, MACMO operates in a weak-excitation regime where each pump pulse induces only a small molecular coherence, as the pulse energy is constrained to keep the average pump power below the damage threshold of ARHCF. Importantly, the molecular coherence generated by one pulse does not fully dephase before the arrival of the next pulse and is constructively reinforced by subsequent pulses. Through repeated interaction with the pulse train, molecular coherence accumulates until a dynamic equilibrium is reached between excitation and dephasing of the oscillation coherence of gas molecules. As the molecular coherence increases, the SRS threshold is reduced because the Raman gain scales with molecular coherence, enabling efficient Raman frequency conversion with low pump pulse energy.

Significant accumulation of molecular coherence is favored when the pump pulse-to-pulse interval is sufficiently short relative to the molecular dephasing time, allowing residual coherence to persist between successive pulses. Although the dephasing time can be extended to microsecond scales at very low pressures ($\ll$1 bar), this is accompanied by a substantial reduction in the Raman gain coefficient, limiting efficient Raman frequency conversion. Realizing MACMO therefore requires a molecular gas at appropriate pressure that simultaneously provides a sufficiently long dephasing time and a high Raman gain coefficient.

A representative example is the Q(1) vibrational mode of hydrogen ($H_2$), which at pressures of tens of bar combines a sub-nanosecond dephasing time with a high Raman gain coefficient (see Fig. S1). Under these conditions, gigahertz repetition rates reduce the pulse-to-pulse interval to a timescale comparable to the Q(1) molecular dephasing time of $H_2$, enabling the accumulation of molecular coherence through MACMO. In addition, the Q(1) vibrational mode of $H_2$ provides a large Raman frequency shift of 4155 cm$^{-1}$ [41].

In this work, the MACMO mechanism is demonstrated in gas-filled ARHCF pumped by a 1 μm ultrafast laser at gigahertz repetition rates. As the pulse train propagates through the fiber, molecular coherence accumulates and drives Raman frequency conversion, generating a Raman pulse train at the fiber output with the same repetition rate.

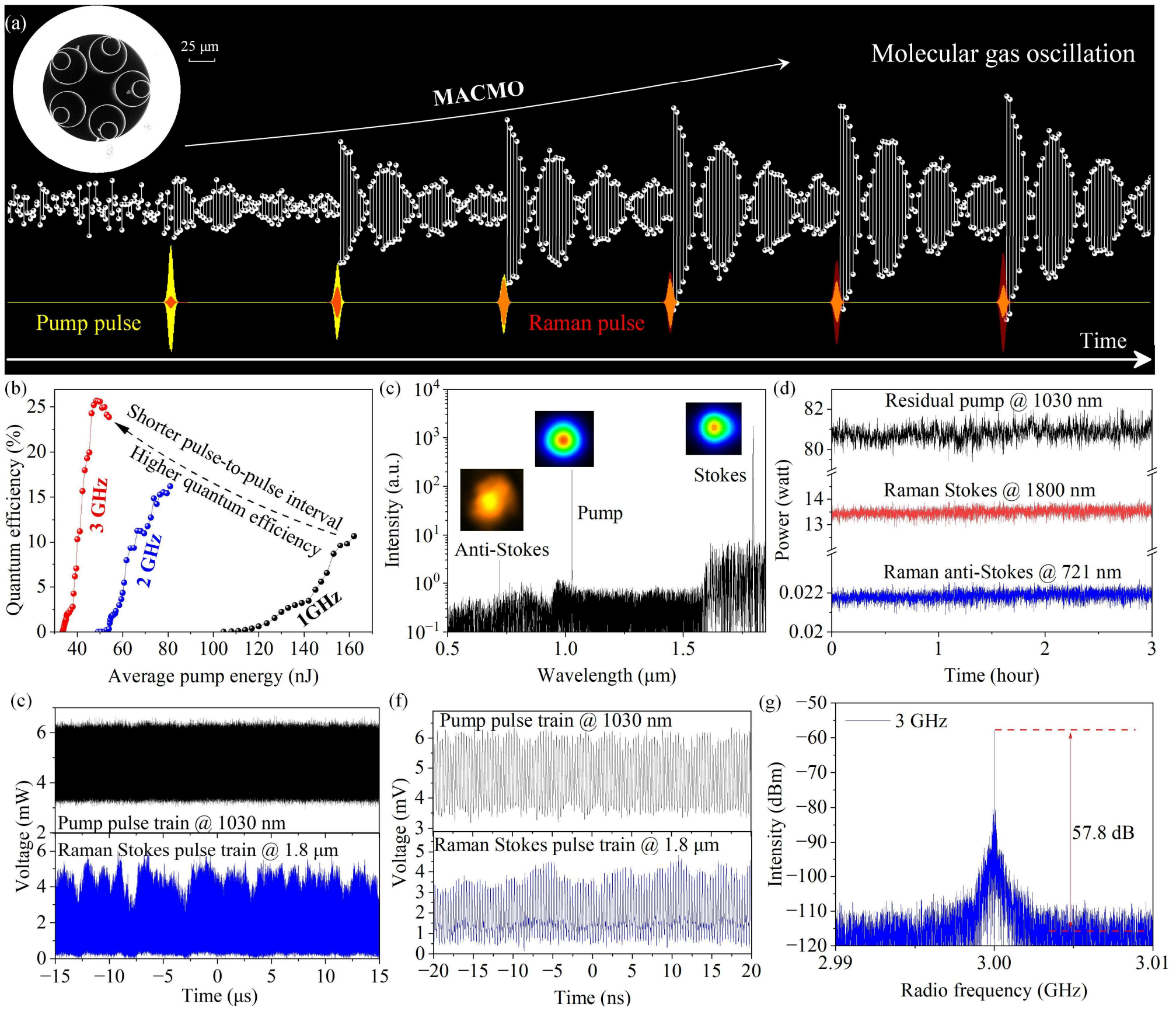

Fig. 1. Concept and validation of MACMO. (a) Schematic illustration of the MACMO mechanism. Inset of (a) shows the cross-section of ARHCF-1 used for MACMO validation in (b-g). (b-g) Experimental performance of MACMO-enabled Raman frequency conversion in a 20-bar $H_2$-filled ARHCF-1 pumped by a 1030-nm pulse train at 1–3 GHz repetition rate. (b) Quantum efficiency of the 1.8 μm vibrational Raman Stokes emission as a function of pump pulse energy at different repetition rates. The raw data for (b) are provided in Fig. S2. The maximum pulse energy is limited by the average power of the 1030 nm pump laser and repetition rate. (c) Optical spectrum at 3 GHz, with corresponding beam profiles of the generated spectral lines. (d) Long-term power stability after a 3-hour system warm-up: residual pump (top), 1.8 μm Raman Stokes (middle), and 721 nm Raman anti-Stokes (bottom), monitored over 3 hours. (e) Pulse trains at 3 GHz repetition rate: Raman Stokes (top) at 7.3 nJ pulse energy and corresponding pump (bottom) at 49 nJ pulse energy. The Raman pulses are delayed by ~30 ns relative to the pump due to optical-path difference. (f) Temporal zoom of the pulse train shown in (e). (g) Radio-frequency spectrum of the 1.8 μm Raman Stokes pulse train at 3 GHz and 7.3 nJ pulse energy, showing the fundamental frequency component of the pulse train. The signal peak to the average noise floor is used for estimating the signal-to-noise ratio. The Raman Stokes power in (d) includes an overall loss of ~40% from lenses, filters, and windows in the experimental setup (Section 1 of Supplement 1); the quantum efficiency in (b) is therefore calculated using the corrected output power. The residual pump signal in (d) exhibits a delay of a few seconds over the two Raman signals because the power meters are not synchronized.

## 3. Validation of MACMO

We experimentally validate MACMO-based Raman frequency conversion using 1st order Q(1) vibrational Raman Stokes generation in $H_2$-filled ARHCF-1 (see Section 1 & 2 of Supplement 1). The pump wavelength is set to 1030 nm, corresponding to a Raman Stokes wavelength of 1.8 μm via the Q(1) frequency shift of $H_2$. The linearly polarized pump delivers a continuous pulse train at 1–3 GHz repetition rates, with a maximum average power of 250 W, corresponding to pulse energies of 83 nJ at 3 GHz and 249 nJ at 1 GHz. The pulse duration is ~30 ps. The $H_2$-filled ARHCF-1 is water-cooled (see Section 2 of Supplement 1) to mitigate heat generated primarily by the Raman quantum defect, which can affect not only long-term operational stability but also the molecular coherence dephasing time through temperature-dependent changes in the Raman linewidth [41]. Details of the experimental setup and characterization methods are provided in Sections 1, 3, and 4 of Supplement 1.

The experiment is conducted at an $H_2$ pressure of 20 bar, corresponding to a molecular dephasing time of ~140 ps for the Q(1) vibrational mode at room temperature. We examine the quantum efficiency of the 1.8 μm Raman Stokes emission as a function of pump pulse energy at different repetition rates. As shown in Fig. 1(b), the quantum efficiency critically depends on repetition rate: increasing the repetition rate reduces the pulse-to-pulse interval, enhances the accumulation of molecular coherence, and leads to a substantial increase in quantum efficiency, which is consistent with the MACMO mechanism. Correspondingly, the SRS threshold substantially decreases from ~104 nJ at 1 GHz to ~34 nJ at 3 GHz repetition rate. For the measurements presented in Fig. 1(b), each data point is recorded after a stabilization period of approximately 5 min to allow the gas-filled ARHCF-1 system to approach thermal equilibrium (see Section 2 of Supplement 1).

A maximum quantum efficiency of 26% is achieved at 3 GHz with a pump pulse energy of 48 nJ (144 W average power), corresponding to a Raman pulse energy of 7.3 nJ and an average output power of ~22 W, with an estimated ~16 W of heat generated by the Raman quantum defect. The measured optical spectrum (Fig. 1(c)) shows frequency conversion of the pump to the vibrational Raman Stokes line. In addition, the 1$^{st}$ order vibrational Raman anti-Stokes emission at 721 nm is clearly observed, demonstrating the potential of MACMO to extend frequency conversion toward the (ultra)violet spectral regions [42,43], for example, by shifting the pump wavelength to 515 nm [44] . The average power of the 721 nm Raman anti-Stokes line is measured to be only ~22 mW due to phase mismatch. Long-term stability is demonstrated by three-hour power monitoring of the Raman Stokes, Raman anti-Stokes, and residual pump signals (Fig. 1(d)), confirming stable operation despite substantial heat generation. The observed stability suggests that the water-cooled ARHCF system can effectively manage the substantial thermal load associated with continuous high-power Raman conversion.

We next examine the temporal characteristics of the Raman output (Fig. 1(e–g)). At a repetition rate of 3 GHz, the Raman laser presents as a continuous pulse train with the same repetition rate as the pump (Fig. 1(e)). Because SRS is seeded by quantum noise, the Raman signal exhibits larger amplitude fluctuations than the pump pulses [45]. However, these fluctuations can be reduced with increasing Raman quantum efficiency (Fig. S3), in agreement with previous observations [45].

Reducing the repetition rate to 1 GHz leads to random discontinuities in the Raman pulse train (Fig. S4). This behavior arises because the 1 ns pulse-to-pulse interval significantly exceeds the molecular dephasing time (~140 ps), preventing stable accumulation of molecular coherence across successive pulses. These fluctuations are therefore expected to be further suppressed at higher repetition rates. In addition, the Raman amplitude stability is influenced by fluctuations in the pump pulse train.

## 4. Investigation of MACMO dynamics

In this section, the dynamics of MACMO are investigated numerically in Section 4.1 and experimentally in Section 4.2. To investigate these dynamics, a GHz-repetition-rate Yb-doped fiber laser operating in burst mode is developed, delivering pump pulses at 1062 nm with a duration of ~23 ps (see Section 5 of Supplement 1). The burst repetition rate is 1 kHz. A 5-m-long ARHCF-2 is selected because its transmission characteristics support first-order $H_2$ Q(1) vibrational Raman Stokes generation at 1.9 μm [46]. The numerical simulations are performed using parameters selected with reference to the pump laser and $H_2$-filled ARHCF-2.

### 4.1 Numerical investigation of MACMO dynamics

We model the 1$^{st}$ order Q(1) vibrational Raman Stokes generation at 1.9 μm in a 5 m long

$H_2$-filled ARHCF-2 at 16 bar pressure. The simulations are based on one-dimensional coupled Maxwell-Bloch equations (Section 6 of Supplement 1). The pump is modeled as a burst of identical Gaussian pulses at 1062 nm, with a pulse duration of 23 ps and a fixed pulse energy of 100 nJ. For a fixed burst duration of 20 ns, this corresponds to 20–60 pulses at repetition rates from 1 to 3 GHz.

Despite the moderate pump pulse energy, the simulations predict efficient frequency conversion. As shown in Fig. 2(a), from ~4 m of propagation, partial pump pulses begin to undergo efficient Raman conversion. The corresponding evolution of quantum efficiency (Fig. 2(b)) reveals a rapid increase along the pulse train: the 1st pulse exhibits a negligible efficiency of $2.2 \times 10^{-5}$ %, which rises to 4.1% for the 7th pulse and reaches 75.2% for the 60th pulse.

This significant efficiency increase originates from molecular coherence accumulated through MACMO. As shown in Fig. 2(c), each pump pulse induces a weak coherence that persists between pulses and is further accumulated by subsequent excitation. The coherence increases by several orders of magnitude over successive pulses and eventually reaches a dynamic steady state. For example, at the fiber output (Fig. 2(d)), the coherence increases from $7.9 \times 10^{-7}$ after the 1st pulse to $6.4 \times 10^{-6}$, $2.7 \times 10^{-5}$, and $1.34 \times 10^{-3}$ after the 2nd, 3rd, and 10th pulses, respectively. This pulse-by-pulse accumulation process enables efficient Raman conversion for later pulses within the burst. Once sufficient molecular coherence is established, subsequent pump pulses are frequency converted in a quasi–steady-state SRS regime. This physical process also makes the Raman pulse duration closely comparable to that of the pump (~23 ps) (Fig. 2(e)), due to suppression of transient SRS regime. Further evolution details are provided in Visualization 1.

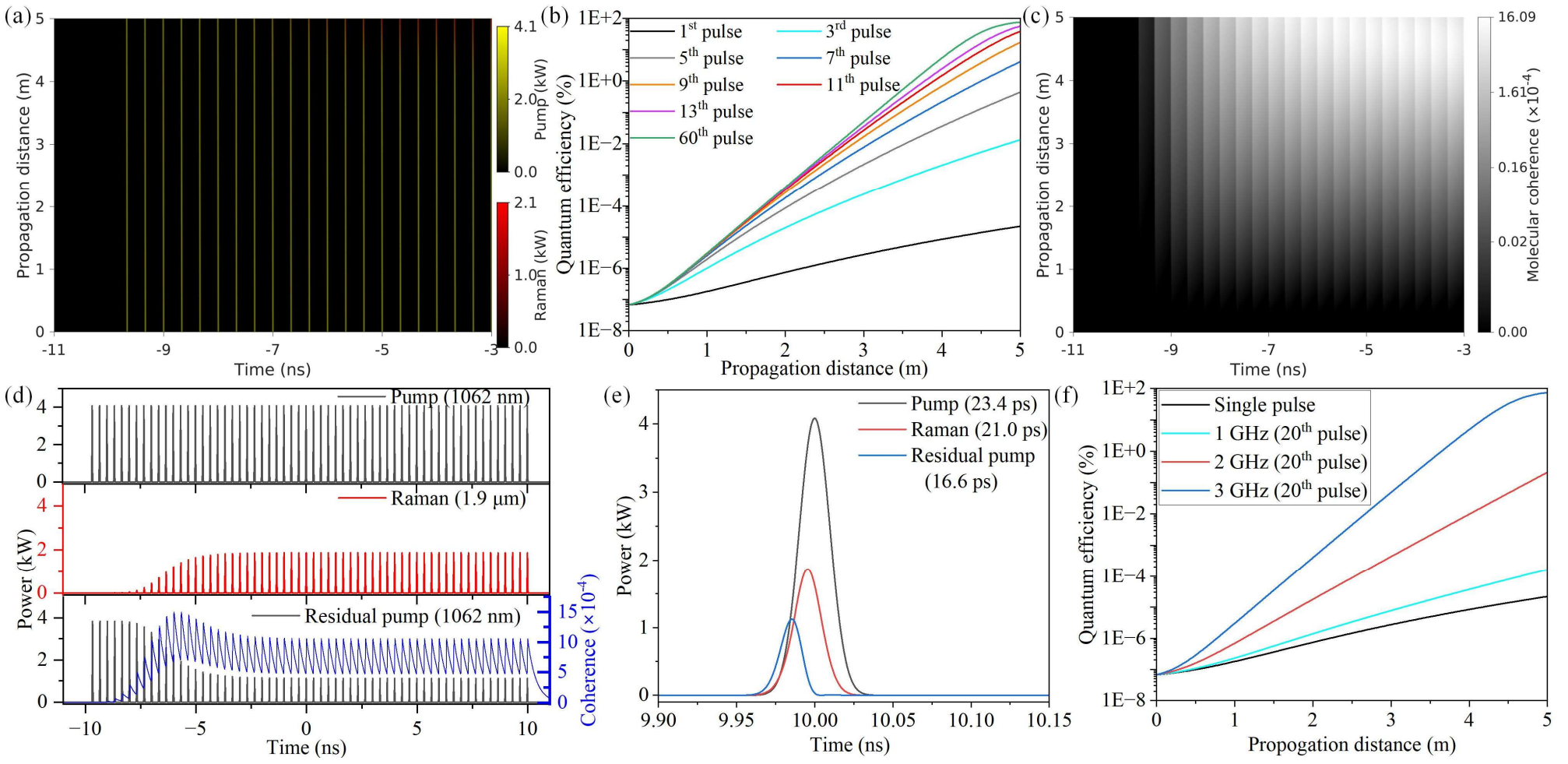

Fig. 2. Numerical modeling of MACMO through 1st order Q(1) vibrational Raman Stokes generation of $H_2$ at 1.9 µm. Simulations were performed in a 5-m-long ARHCF-2 filled with H2 at 16 bar. (a) Evolution of pump pulses' power (yellow) and frequency-converted Raman pulses' power (red) along the fiber. The simulation uses a burst of 60 ultrafast pump pulses at 3 GHz repetition rate and 1062 nm wavelength; for clarity, only the first 21 pulses are shown. (b) Raman quantum efficiencies of selected pump pulses (1st, 3rd, 5th, 7th, 9th, 11th, 13th, and 60th) within the burst. (c) Evolution of hydrogen molecular coherence in logarithmic scale, corresponding to (a). (d) Temporal profiles of the full pump burst, generated Raman burst, residual pump, and hydrogen molecular coherence at the fiber output. (e) Temporal profiles of the last (60th) pump pulse, its corresponding Raman pulse, and the residual pump pulse. (f) Quantum efficiencies of Raman Stokes generation from the 20th pump pulse at repetition rates of 1 GHz, 2 GHz, and 3 GHz. For comparison, the efficiency of a single pump pulse is also included. In this simulation, each pump pulse has 100 nJ energy and a Gaussian temporal profile with 23 ps pulse duration at 1062 nm wavelength. The simulated results in (a) and (c) over the complete time window of 32 ns are provided in Fig. S5.

Consistent with Fig. 1(b), the simulated conversion efficiency strongly depends on repetition rate. Reducing the repetition rate weakens coherence accumulation via MACMO and leads to a substantial decrease in efficiency. For example, lowering the repetition rate from 3

GHz to 1 GHz reduces the maximum efficiency of the 20th pulse from 74% to $1.6 \times 10^{-4}$ % (Fig. 2(f)), as the 1 ns pulse-to-pulse interval far exceeds the 375 ps dephasing time of the Q(1) mode at 16 bar. In the single-pulse limit, the efficiency is only $2.2 \times 10^{-5}$ %.

## 4.2 Experimental investigation of MACMO dynamics

The MACMO dynamics revealed in Fig. 2(d) are experimentally validated by pumping $H_2$-filled ARHCF-2 with the 1062 nm laser operating in burst mode. Under these conditions, 1st order vibrational Raman Stokes emission at 1.9 μm is generated. Figure 3(a) shows the measured temporal profiles at an average pump energy of 678 nJ and repetition rate of 2 GHz. Consistent with the simulations, the residual pump indicates that pulses at the leading edge of the burst are not significantly depleted. As a result, the Raman pulse train emerges with a delay of ~5 ns relative to the pump burst, reflecting the accumulation of molecular coherence via MACMO. Additional details are provided in Visualization 2. The number of undepleted pulses at the leading edge of the burst depends on the gas pressure, as the molecular response can accumulate over successive pulses, giving rise to transient Raman dynamics on the nanosecond timescale (Fig. S6). The measured autocorrelation traces (Fig. 3(b)) show that both the pump and Raman pulses have pulse durations of ~23 ps, in good agreement with simulations and confirming suppression of the transient SRS regime. Further details are provided in Section 7 of Supplement 1.

In addition to $H_2$, we investigate MACMO using $CO_2$ and $CH_4$ as Raman-active media in ARHCF-2. These gases are widely used for Raman frequency conversion [47–49]. The $v_1/2v_2$ vibrational mode of $CO_2$ provides a long molecular dephasing time (~1 ns at a few bar; Fig. S1), enabling MACMO at gigahertz repetition rates. Accordingly, at 3 GHz repetition rate and 478 nJ average pulse energy, the pump is converted to the 1st order Raman Stokes at 1.25 μm with a quantum efficiency of 38% (Fig. S7(e)). In contrast, $CH_4$ has a much shorter vibrational dephasing time (~30 ps under few bar pressure), which is insufficient to sustain coherence buildup between successive pulses. Consequently, no Raman signal is observed in the 1.5 μm region under the same pump conditions. For comparison, Fig. 3(c) shows the measured output spectra obtained using $H_2$, $CO_2$, and $CH_4$. The different responses of $H_2$, $CO_2$ and $CH_4$ directly support the MACMO mechanism, demonstrating that efficient frequency conversion occurs only when the molecular dephasing time is sufficiently long or comparable to the pulse-to-pulse interval to enable molecular coherence accumulation across successive pulses.

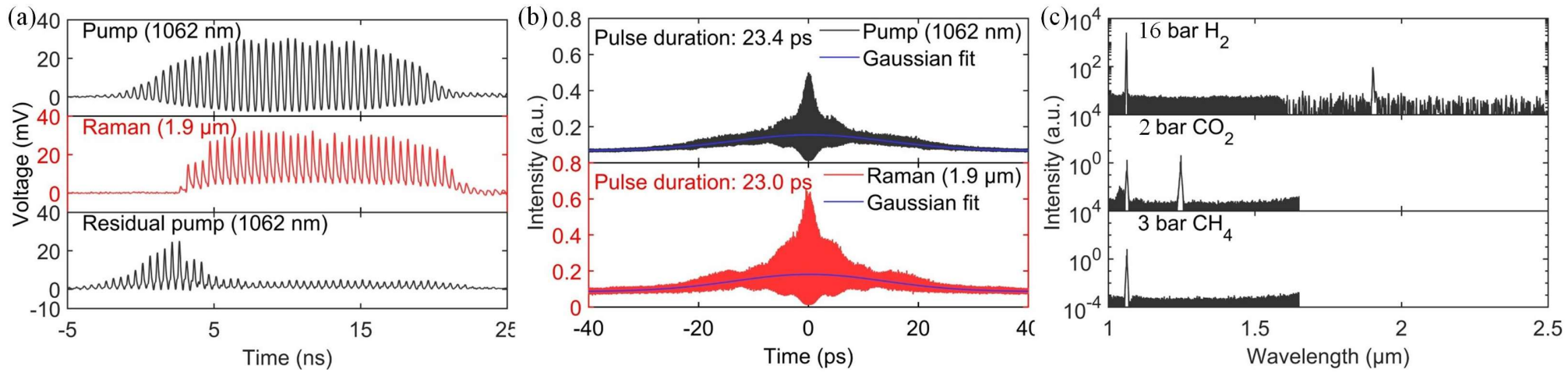

Fig. 3. Experimental validation of MACMO dynamics in gas-filled ARHCFs pumped by a burst-mode pulse train at 2 GHz repetition rate. (a) Measured temporal profiles of a pump burst (top), its corresponding Raman Stokes burst (middle), and residual pump burst (bottom) output from $H_2$-filled ARHCF-2. (b) Corresponding autocorrelation traces and Gaussian fits of the pump and Raman pulses. Autocorrelation traces over a time window of 1.1 ns are provided in Fig. S7(a). (c) Measured optical spectra from $H_2$-filled ARHCF-2 (top), $CO_2$-filled ARHCF-2 (middle), and $CH_4$-filled ARHCF-2 (bottom). The gas pressures for $H_2$, $CO_2$, and $CH_4$ are 16 bar, 2 bar, and 3 bar, respectively. Further details regarding this section are provided in Section 7 of Supplement 1.

## 5. Discussion & outlook

While MACMO-driven frequency conversion is demonstrated in this work, the broader implications of coherent molecular oscillations remain largely unexplored both in physics and applications. Here we discuss future directions in terms of wavelength accessibility, pulse duration, and gas pressure.

The wavelength accessibility of MACMO is central to its relevance across a broad range of applications. For example, mid- and far-infrared atmospheric transmission windows are advantageous for Earth–satellite free-space optical communication [17,50], while laser micromachining performance strongly depends on the operating wavelength [14,51]. The MACMO mechanism inherits the spectral flexibility of conventional gas-filled ARHCF Raman lasers, which have been demonstrated from the deep ultraviolet to the mid-infrared, including broadband tunability in the near- and mid-infrared [52]. In Section 8 of Supplement 1, we further demonstrate the wavelength scalability of MACMO (in burst mode), including tunable mid-infrared Raman Stokes generation through cascaded $CO_2$-filled ARHCF-3 and $H_2$-filled ARHCF-4, as well as visible anti-Stokes emission (737 nm and 564 nm) in $H_2$-filled ARHCF-2.

A key practical question is how MACMO performs as the pump pulse duration is reduced below the ~30 ps regime explored here. As the pulse duration decreases, the quantum efficiency is expected to be reduced due to increased contribution from the transient SRS regime, spectral broadening of the pump, and dispersion-induced walk-off in the fiber. In particular, the group-velocity difference between the pump and Raman pulses can cause them to walk off temporally during propagation (see Fig. 2e), reducing their temporal overlap and eventually suppressing Raman frequency conversion. Minimizing this group-velocity mismatch is therefore favorable for efficient MACMO-based Raman frequency conversion, particularly for shorter pump pulses. This can be achieved by engineering the dispersion profile of the hollow-core fiber, for example, through appropriate selection of its core diameter, cladding geometry, and gas pressure to reduce the group-velocity difference between the pump and Raman wavelengths.

In the femtosecond regime, impulsive SRS becomes dominant [53]. In this limit, molecular coherence evolves with a well-defined phase, making its interaction with successive pulses highly phase sensitive [35]. Constructive accumulation therefore requires precise synchronization between the pulse train and molecular oscillations, whereas even small frequency mismatches lead to cumulative phase slippage and degradation of coherence over time [36,54]. By contrast, as demonstrated in this work, MACMO operates robustly in the transient/quasi-steady Raman regime, where phase sensitivity is effectively averaged out. Nevertheless, extending MACMO into the impulsive regime may provide deeper insight into its underlying dynamics and enable significantly higher molecular coherence, potentially approaching its theoretical maximum of 0.5.

Gas pressure also plays a critical role in MACMO. Increasing the pressure enhances the Raman gain coefficient but simultaneously shortens the molecular dephasing time (Fig. S1), leading to competing effects. Reduced coherence lifetime weakens multi-pulse accumulation, whereas the suppression of the transient Raman regime is an advantageous factor. Optimal performance therefore requires balancing Raman gain, transient response, and MACMO (see Section 7 of Supplement 1). This interplay also implies the potential of extending MACMO to repetition rates well beyond the ~3 GHz explored here. At higher gas pressures, enhanced Raman gain can compensate for the shorter dephasing time when combined with reduced pulse-to-pulse intervals, potentially enabling operation at tens to hundreds of gigahertz repetition rates. Such high repetition rates combined with high pulse energies are currently accessible only within limited regions of the near-infrared spectrum [55,56].

**Funding.** LUNDBECK Fonden (Grant No. R276-2018-869); VILLUM Fonden (Grant No. 36063, Grant No. 40964).

**Disclosures.** The authors declare no conflicts of interest.

**Data availability statement.** All data generated in this work have been deposited in Figshare (https://doi.org/10.6084/m9.figshare.34333140). The dataset is currently under embargo and will be made publicly available upon acceptance of the manuscript for publication.

**Supplementary Document.** See Supplement 1 for supporting content.

**Author contributions.** Yazhou Wang conceived the MACMO concept, conducted the numerical simulations and experimental investigation of the MACMO-based frequency conversion. Christos Markos contributed to discussions related to MACMO physics and mechanism, and financially supported the experimental validation of the concept. Timothy Bate, J. E. Antonio-Lopez, and Rodrigo Amezcua-Correa fabricated and characterized ARHCF-1, ARHCF-2, and ARHCF-3. Ian Davidson, Qiang Fu, Jaroslaw Rzegocki, Gregory T. Jasion, and Francesco Poletti fabricated and characterized ARHCF-4. Yazhou Wang, and Christos Markos, and Marcello Meneghetti, prepared the figures and supplementary videos. Yazhou Wang and Christos Markos prepared the initial manuscript draft and all authors contributed to reviewing and revising the manuscript.

## References

1. A. Barh, B. Ö. Alaydin, J. Heidrich, M. Gaulke, M. Golling, C. R. Phillips, and U. Keller, " High-power low-noise 2-GHz femtosecond laser oscillator at 2.4 μm," Opt. Express 30, 5019–5025 (2022).
2. D. Kazakov, T. P. Letsou, M. Piccardo, L. L. Columbo, M. Brambilla, F. Prati, S. Dal Cin, M. Beiser, N. Opačak, P. Ratra, M. Pushkarsky, D. Caffey, T. Day, L. A. Lugiato, B. Schwarz, and F. Capasso, "Driven bright solitons on a mid-infrared laser chip," Nature 641, 83–89 (2025).
3. R. Rockmore, A. Laurain, J V Moloney, and R. Jason Jones, "Offset-free mid-infrared frequency comb based on a mode-locked semiconductor laser," Opt. Lett. 44, 1797–1800 (2019).
4. N. Hoghooghi, S. Xing, P. Chang, D. Lesko, A. Lind, G. Rieker, and S. Diddams, "Broadband 1-GHz mid-infrared frequency comb," Light Sci. Appl. 11, 264 (2022).
5. C. Li, H. Xuan, L. Winkelmann, and I. Hartl, "3GHz, 257nm picosecond source for electron guns," In Proc. 8th EPS-QEOD Europhoton Conf. (Barcelona, Spain, 2018).
6. T.-H. Wu, L. Ledezma, C. Fredrick, P. Sekhar, R. Sekine, Q. Guo, R. M. Briggs, A. Marandi, and S. A. Diddams, "Visible-to-ultraviolet frequency comb generation in lithium niobate nanophotonic waveguides," Nat. Photonics 18, 218–223 (2024).
7. C. P. Bauer, L. Lang, B. Willenberg, J. M. Pereira, J. Pupeikis, and U. Keller, "Ultrafast 1-GHz repetition rate dual-comb optical parametric oscillator," Opt. Express 33, 39130–39139 (2025).
8. Z. Zhao, B. M. Dunham, and F. W. Wise, "Generation of 167 W infrared and 124 W green power from a 1.3-GHz, 1-ps rod fiber amplifier," Opt. Express 22, 25065–25070 (2014).
9. J. Lim, H.-W. Chen, S. Xu, Z. Yang, G. Chang, and F. X. Kärtner, "3 GHz, watt-level femtosecond Raman soliton source," Opt. Lett. 39, 2060–2063 (2014).
10. J. W. Nicholson, R. Ahmad, A. DeSantolo, and Z. Várallyay, "High average power, 10 GHz pulses from a very-large-mode-area, Er-doped fiber amplifier," J. Opt. Soc. Am. B 34, A1–A6 (2017).
11. N. Hoghooghi, P. Chang, S. Egbert, M. Burch, R. Shaik, S. A. Diddams, P. Lynch, and G. B. Rieker, "GHz repetition rate mid-infrared frequency comb spectroscopy of fast chemical reactions," Optica 11, 876–882 (2024).
12. S.-W. Chu, T.-M. Liu, C.-K. Sun, C.-Y. Lin, and H.-J. Tsai, "Real-time second-harmonic-generation microscopy based on a 2-GHz repetition rate Ti:sapphire laser," Opt. Express 11, 933–938 (2003).
13. N. Ji, J. C. Magee, and E. Betzig, "High-speed, low-photodamage nonlinear imaging using passive pulse splitters," Nat. Methods 5, 197–202 (2008).
14. C. Kerse, H. Kalaycıoğlu, P. Elahi, B. Çetin, D. K. Kesim, Ö. Akçaalan, S. Yavaş, M. D. Aşık, B. Öktem, H. Hoogland, R. Holzwarth, and F. Ö. Ilday, "Ablation-cooled material removal with ultrafast bursts of pulses," Nature 537, 84–88 (2016).
15. Z. Chen, C. Meng, H. Lin, D. Fixler, and H. Zhang, "Spatially confined tumor phototherapy enabled by GHz laser thermal accumulation dynamics," Adv. Photonics Nexus 4, 066009 (2025).
16. K. Mishchik, G. Bonamis, J. Qiao, J. Lopez, E. Audouard, E. Mottay, C. Hönninger, and I. Manek-Hönninger, "High-efficiency femtosecond ablation of silicon with GHz repetition rate laser source," Opt. Lett. 44, 2193–2196 (2019).
17. O. Spitz, P. Didier, L. Durupt, D. A. Díaz-Thomas, A. N. Baranov, L. Cerutti, and F. Grillot, "Free-space communication with directly modulated mid-infrared quantum cascade devices," IEEE J. Sel. Top. Quantum Electron. 28, 1–9 (2022).
18. H. Xu, J. Xu, X. Li, L. Feng, S. Huang, and J. Li, "High power drive laser system for photocathode at IHEP," Opt. Express 29, 29550–29556 (2021).
19. M. Piotrowski, M. Alessandro Medina, M. Schellhorn, G. Spindler, and A. Hildenbrand-Dhollande, "Effects of pump pulse energy and repetition rate on beam quality in a high-power mid-infrared $ZnGeP_2$ OPO," Opt. Express 29, 2577–2586 (2021).
20. H. Carstens, M. Högner, T. Saule, S. Holzberger, N. Lilienfein, A. Guggenmos, C. Jocher, T. Eidam, D. Esser, V. Tosa, V. Pervak, J. Limpert, A. Tünnermann, U. Kleineberg, F. Krausz, and I. Pupeza, "High-harmonic generation at 250 MHz with photon energies exceeding 100 eV," Optica 3, 366–369 (2016).
21. I. Pupeza, C. Zhang, M. Högner, and J. Ye, "Extreme-ultraviolet frequency combs for precision metrology and attosecond science," Nat. Photonics 15, 175–186 (2021).
22. S. Gao, Y. Wang, W. Ding, D. Jiang, S. Gu, X. Zhang, and P. Wang, "Hollow-core conjoined-tube negative-

curvature fibre with ultralow loss," Nat. Commun. 9, 2828 (2018).
23. A. I. Adamu, Md. S. Habib, C. R. Petersen, J. E. A. Lopez, B. Zhou, A. Schülzgen, M. Bache, R. Amezcua-Correa, O. Bang, and C. Markos, "Deep-UV to mid-ir supercontinuum generation driven by mid-IR ultrashort pulses in a gas-filled hollow-core fiber," Sci. Rep. 9, 4446 (2019).
24. M. Petrovich, E. Numkam Fokoua, Y. Chen, H. Sakr, A. I. Adamu, R. Hassan, D. Wu, R. Fatobene Ando, A. Papadimopoulos, S. R. Sandoghchi, G. Jasion, and F. Poletti, "Broadband optical fibre with an attenuation lower than 0.1 decibel per kilometre," Nat. Photonics (2025).
25. B. Plosz, A. Lekosiotis, M. Sabbah, F. Belli, C. Brahms, and J. C. Travers, "Supercontinuum generation in a methane-filled hollow-core antiresonant fiber," Opt. Lett. 50, 1285–1288 (2025).
26. M. A. Cooper, J. Wahlen, S. Yerolatsitis, D. Cruz-Delgado, D. Parra, B. Tanner, P. Ahmadi, O. Jones, Md. S. Habib, I. Divliansky, J. E. Antonio-Lopez, A. Schülzgen, and R. Amezcua Correa, "2.2 kW single-mode narrow-linewidth laser delivery through a hollow-core fiber," Optica 10, 1253–1259 (2023).
27. H. C. H. Mulvad, S. Abokhamis Mousavi, V. Zuba, L. Xu, H. Sakr, T. D. Bradley, J. R. Hayes, G. T. Jasion, E. Numkam Fokoua, A. Taranta, S.-U. Alam, D. J. Richardson, and F. Poletti, "Kilowatt-average-power single-mode laser light transmission over kilometre-scale hollow-core fibre," Nat. Photonics 16, 448–453 (2022).
28. J. Shi, B. Rao, Z. Chen, Z. Wang, G. Sun, Z. Xu, Z. Huang, P. Li, Z. Dong, M. Fu, X. Tian, B. Yang, J. Zhang, Z. Zhou, T. Li, L. Zhang, B. Shui, C. Gao, and J. Chen, "All-fiber highly efficient delivery of 2 kW laser over 2.45 km hollow-core fiber," Nat. Commun. 16, 8965 (2025).
29. X. Zhu, F. Yu, D. Wu, S. Chen, Y. Jiang, and L. Hu, "Laser-induced damage of an anti-resonant hollow-core fiber for high-power laser delivery at 1 μm," Opt. Lett. 47, 3548–3551 (2022).
30. A. Lekosiotis, F. Belli, C. Brahms, M. Sabbah, H. Sakr, I. A. Davidson, F. Poletti, and J. C. Travers, "On-target delivery of intense ultrafast laser pulses through hollow-core anti-resonant fibers," Opt. Express 31, 30227–30238 (2023).
31. D. E. Couch, D. D. Hickstein, D. G. Winters, S. J. Backus, M. S. Kirchner, S. R. Domingue, J. J. Ramirez, C. G. Durfee, M. M. Murnane, and H. C. Kapteyn, "Ultrafast 1 MHz vacuum-ultraviolet source via highly cascaded harmonic generation in negative-curvature hollow-core fibers," Optica 7, 832–837 (2020).
32. A. Deng, T. Gavara, M. R. A. Hassan, D. Xiong, M. I. Hasan, and W. Chang, "Microjoule-level mid-infrared femtosecond pulse generation in hollow-core fibers," Laser Photon. Rev. 2200882 (2023).
33. W. Song, Q. Sui, Q. Zhang, X. Zhang, D. Hong, S. Qiao, and P. Wang, "10 W-level high-energy $CH_4$-filled hollow-core fiber picosecond Raman laser at 2.8 μm," J. Lightwave Technol. 43, 9375–9381 (2025).
34. L. L. Losev, V. S. Pazyuk, and A. V Gladyshev, "Femtosecond hydrogen Raman frequency-shifter/pulse-compressor based on revolver fiber," IEEE J. Sel. Top. Quantum Electron. 30, 1–5 (2024).
35. S. Baker, I. A. Walmsley, J. W. G. Tisch, and J. P. Marangos, "Femtosecond to attosecond light pulses from a molecular modulator," Nat. Photonics 5, 664–671 (2011).
36. F. Belli, A. Abdolvand, J. C. Travers, and P. St. J. Russell, "Control of ultrafast pulses in a hydrogen-filled hollow-core photonic-crystal fiber by Raman coherence," Phys. Rev. A (Coll. Park). 97, 13814 (2018).
37. M. Jain, H. Xia, G. Y. Yin, A. J. Merriam, and S. E. Harris, "Efficient nonlinear frequency conversion with maximal atomic coherence," Phys. Rev. Lett. 77, 4326–4329 (1996).
38. G. Korn, O. Dühr, and A. Nazarkin, "Observation of Raman self-conversion of fs-pulse frequency due to impulsive excitation of molecular vibrations," Phys. Rev. Lett. 81, 1215–1218 (1998).
39. C. Vicario, M. Shalaby, A. Konyashchenko, L. Losev, and C. P. Hauri, "High-power femtosecond Raman frequency shifter," Opt. Lett. 41, 4719–4722 (2016).
40. R. Tyumenev, J. Hammer, N. Y. Joly, P. St. J. Russell, and D. Novoa, "Tunable and state-preserving frequency conversion of single photons in hydrogen," Science (1979). 376, 621–624 (2022).
41. W. K. Bischel and M. J. Dyer, "Wavelength dependence of the absolute Raman gain coefficient for the Q(1) transition in $H_2$," J. Opt. Soc. Am. B 3, 677–682 (1986).
42. M. K. Mridha, D. Novoa, S. T. Bauerschmidt, A. Abdolvand, and P. St.J. Russell, "Generation of a vacuum ultraviolet to visible Raman frequency comb in $H_2$-filled kagomé photonic crystal fiber," Opt. Lett. 41, 2811–2814 (2016).
43. R. Mears, K. Harrington, J. C. Knight, W. J. Wadsworth, J. M. Stone, and T. A. Birks, "Perspective: hollow core optical fibres for ultraviolet and visible wavelengths," Adv. Sci., e17476 (2026).
44. Z. Zhao, B. M. Dunham, and F. W. Wise, "Generation of 167 W infrared and 124 W green power from a 1.3-GHz, 1-ps rod fiber amplifier," Opt. Express 22, 25065–25070 (2014).
45. Y. Wang, A. I. Adamu, M. K. Dasa, J. E. Antonio-Lopez, M. S. Habib, R. Amezcua-Correa, O. Bang, and C. Markos, "Noise performance and long-term stability of near- and mid-IR gas-filled fiber Raman lasers," J. Light. Technol. 39, 3560–3567 (2021).
46. A. I. Adamu, Y. Wang, M. S. Habib, M. K. Dasa, J. E. Antonio-Lopez, R. Amezcua-Correa, O. Bang, and C. Markos, "Multi-wavelength high-energy gas-filled fiber Raman laser spanning from 1.53 μm to 2.4 μm," Opt. Lett. 46(3), 452–455 (2021).
47. Y. Wang, L. Hong, C. Zhang, J. Wahlen, J. E. Antonio-Lopez, M. K. Dasa, A. I. Adamu, R. Amezcua-Correa, and C. Markos, "Synthesizing gas-filled anti-resonant hollow-core fiber Raman lines enables access to the molecular fingerprint region," Nat. Commun. 15, 9427 (2024).
48. Z. Li, W. Huang, Y. Cui, and Z. Wang, "Efficient mid-infrared cascade Raman source in methane-filled hollow-core fibers operating at 2.8 μm," Opt. Lett. 43, 4671–4674 (2018).
49. M. S. Astapovich, A. V Gladyshev, M. M. Khudyakov, A. F. Kosolapov, M. E. Likhachev, and I. A. Bufetov, "

Watt-level nanosecond 4.42 μm Raman laser based on silica fiber," IEEE Photonics Technol. Lett. 31, 78–81 (2019).
50. L. Flannigan, L. Yoell, and C. Xu, "Mid-wave and long-wave infrared transmitters and detectors for optical satellite communications—a review," J. Opt. 24, 043002 (2022).
51. X. Liu, D. Du, and G. Mourou, "Laser ablation and micromachining with ultrashort laser pulses," IEEE J. Quantum Electron. 33, 1706–1716 (1997).
52. Y. Wang, et al. "Tunable, high pulse energy and narrow linewidth gas-filled fiber laser across near- and mid-infrared," Laser & Photonics Rev. e71703 (2026).
53. N. Zhavoronkov and G. Korn, "Generation of single intense short optical pulses by ultrafast molecular phase modulation," Phys. Rev. Lett. 88, 203901 (2002).
54. A. M. Weiner, D. E. Leaird, G. P. Wiederrecht, and K. A. Nelson, "Femtosecond pulse sequences used for optical manipulation of molecular motion," Science (1979). 247, 1317–1319 (1990).
55. E. Yoshida and M. Nakazawa, "Low-threshold 115-GHz continuous-wave modulational-instability erbium-doped fiber laser," Opt. Lett. 22, 1409–1411 (1997).
56. Y. D. Gong, P. Shum, D. Y. Tang, C. Lu, and X. Guo, "660GHz soliton source based on modulation instability in a short cavity," Opt. Express 11, 2480–2485 (2003).

# Multi-pulse accumulation of gas molecular coherence enables gigahertz ultrafast frequency conversion

YAZHOU WANG,[1,*] MARCELLO MENEGHETTI,[1] IAN DAVIDSON,[2] TIMOTHY BATE,[3] J. E. ANTONIO-LOPEZ,[3] QIANG FU,[2] JAROSLAW RZEGOCKI,[2] GREGORY T. JASION,[2] RODRIGO AMEZCUA-CORREA,[3] FRANCESCO POLETTI,[2] AND CHRISTOS MARKOS[1,4,*]

[1]*DTU Electro, Technical University of Denmark, 2800 Kgs. Lyngby, Denmark*
[2]*Optoelectronics Research Centre, University of Southampton, Southampton, UK*
[3]*CREOL, The College of Optics and Photonics, University of Central Florida, Orlando, Florida 32816, USA*
[4]*NORBLIS ApS, Virum, Denmark*

**yazwang@dtu.dk; chmar@dtu.dk*

**Section 1. Configuration of high power MACMO-based gas-filled ARHCF Raman laser system**

Methods for preparing gas-filled ARHCFs under low-power operation (e.g., < 1 W) are described in Ref. [1]. Here we detail the configuration used for high-power (> 100 W) MACMO-based frequency conversion based on $H_2$-filled ARHCF-1.

The ARHCF-1 is placed in a stainless-steel tube (outer diameter: 4 mm) for gas filling. This tube is inserted into a larger nylon tube (outer diameter: 12 mm), which enables water flow for efficient thermal management. The stainless-steel tube (and enclosed fiber) has a length of 5 m and is coiled with a diameter of ~40 cm.

Both ends of the stainless-steel tube are sealed with custom gas cells. Each gas cell is equipped with a B-coated flat optical window (WG41050-B, Thorlabs) to minimize back-reflection and prevent optical damage under high-power pumping at 1030 nm. The B-coated windows introduce an ~18% transmission loss at the 1.8 μm Raman Stokes wavelength. The input end of the ARHCF-1 is fixed in a V-groove within the input gas cell, which is cooled via an integrated water channel. The assembled system is tested to ensure a hydrogen leak rate below 3% per day at ~20 bar.

The pump laser described in Section 3 of Supplement 1 is coupled into the ARHCF-1 system using a coupling lens (LA4052-B-ML, Thorlabs) mounted on a precision translation stage (NanoMax 300, Thorlabs). Coupling efficiency is first optimized under ambient air to avoid nonlinear effects. Under these conditions, the coupling efficiency is initially below 90% and increases to ~94% at pump powers above ~100 W due to improved beam quality of the Yb-doped rod-amplifier used in the pump laser (see Section 3 of Supplement 1). During power scaling, the coupling lens is fine-tuned to compensate for thermal lensing effect and mechanical drift due to temperature arising.

After optimization, the ARHCF-1 is filled with 20 bar $H_2$ for MACMO-based Raman frequency conversion. When the Raman laser is generated, maximizing the Raman power is used as an index to optimize coupling efficiency. However, in this case, the Raman power variation is caused by both the coupling efficiency and the thermal effects of Raman quantum defect, imposing an experimental challenge on optimizing the coupling efficiency to ~94% particularly when the Raman Stokes power at 1.8 μm reaches to watt level. To mitigate the thermal damage risk caused by the possible low coupling efficiency, two measures are adopted. First, the 1030 nm pump power is limited to $\lesssim$150 W, despite a maximum available power of 250 W. Second, the pump power is increased incrementally, in steps corresponding to $\lesssim$ 0.5 W Raman output, with each step held for ~5 min to gradually reach thermal equilibrium and then the coupling lens is fine adjusted to maximize the Raman power. This measure also avoids rapid temperature excursions that may induce mechanical failure of the silica fiber.

Before the long-term stability measurements (Fig. 1(d)), the system is warmed up for ~3 hours. During this warming-up period, the coupling lens is occasionally adjusted to re-optimize coupling efficiency by maximizing the Raman output power.

The output beam from the ARHCF-1 is first collimated with a plano-convex lens (LA4148-B-ML, Thorlabs), and then a dichroic mirror (DMLP1180, Thorlabs) is used to separate the 1.8 μm Raman Stokes laser from other beams including residual pump beam at 1030 nm and Raman anti-Stokes at 721 nm. Bottom of Fig. S8 shows the optical path.

i) Raman Stokes beam. The transmitted Raman Stokes beam is spectrally cleaned using a long-pass filter (FELH1600, Thorlabs) and then passes through a wedged window (WW41050, Thorlabs), which reflects 5.3% of the power for diagnostic measurements (temporal profile and beam profile). The main transmitted beam is used for power monitoring. The total insertion loss of the Raman path is ~40%, arising from the B-coated window (18%), B-coated collimation lens (18%), dichroic mirror (4.3%), long-pass filter (0.6%), and wedged window (5.3%).

ii) Residual pump and anti-Stokes beams. The reflected beam from DMLP1180 is further separated using a second dichroic mirror (DMSP900, Thorlabs). The residual pump is reflected to a power meter, while the Raman anti-Stokes beam is transmitted for spectral, power, and beam-profile characterization. Due to non-ideal spectral selectivity of the dichroic optics, a small fraction of pump and Stokes light remains mixed with the anti-Stokes beam, enabling measurement of the complete spectrum shown in Fig. 1(c). For accurate anti-Stokes power measurements, a short-pass filter (FESH05950, Thorlabs) is used to suppress longer wavelengths.

**Section 2: Properties of ARHCFs**

In this work, four ARHCFs are used, and they are numbered as ARHCF-1, ARHCF-2, ARHCF-3, and ARHCF-4. All these ARHCFs are coiled in a bend diameter of ~30-40 cm. Details of these ARHCFs are provided below.

i) Characteristics of the ARHCF-1 are provided in Fig. S9(a) and S9(b). This fiber has a nested cladding structure, forming a hollow-core region with diameter of 25 μm. The thicknesses of both the large capillary and nested capillary are 320 nm and 316 nm, enabling a transmission window of ~ 950-1800 nm, as indicated with Fig. S9(b). The length of ARHCF-1 is 5 m.

ii) ARHCF-2 has a nested cladding, forming a core diameter of 37 μm [2]. Capillary wall thicknesses are ~406 nm and ~630 nm for the large capillary and nested capillary, respectively. The fiber length is 5 m. All simulations presented in this work adopt the structural parameters of this fiber. The finite-element method using the software COMSOL is employed to simulate the loss and group velocity of this ARHCF as a function of wavelength, at three different $H_2$ pressures. Fig. S9(c) shows the geometric model built with COMSOL, and the corresponding mode field distribution at 1062 nm under a bend diameter of 30 cm. Fig. S9(d) and 9(e) shows the simulated group velocity and loss at different $H_2$ pressures. The results indicate that $H_2$ pressure affects group velocity, while its impact on loss is negligible.

iii) ARHCF-3 has a 32.8 μm core diameter surrounded by seven non-nested silica capillaries, where each capillary has a diameter of 16.1 μm and a wall thickness of 323 nm [1]. The fiber length is 5 m.

iv) Characteristics of ARHCF-4 are provided in Fig. S9(f) and S9(g). This fiber has a nested cladding structure, forming a hollow-core region with diameter of 32 μm. The thickness of both the large capillary and nested capillary is 394 nm, enabling a transmission window of ~ 1200-2500 nm, as indicated with Fig. S9(g). The length of ARHCF-4 is 10 m.

**Section 3. Design of 1030 nm high repetition rate pump laser in continuous mode**

The top part of Fig. S8 shows the configuration of the pump laser emitting continuous pulse train at gigahertz repetition rate and 1030 nm wavelength, comprising a high repetition rate seed laser, and an optical amplification module.

The high-repetition-rate seed laser is generated by modulating a single-frequency continuous-wave (CW) laser (TEC-500-1060-030, Sacher Lasertechnik) with a high-speed Mach-Zehnder intensity modulator (NIR-MX-LN-40, Exail Technologies). The modulator features a 40 GHz bandwidth, 10 ps rise/fall time, and an extinction ratio of ~30 dB across 1030-1060 nm range. A modulator bias controller (MBC-DGLAB-A3, Exail) is used for stabilizing the modulator. The CW seed laser wavelength is tunable from 1030 nm to 1090 nm with a linewidth of 10 kHz, while fixed to 1030 nm in this section. The Mach-Zehnder modulator is driven by a voltage pulse generator (EPG-LAB-30ps, Exail), which emits a voltage pulse train with adjustable duration from 30 ps to 130 ps and a tunable repetition rate ranging from single-shot to 3.5 GHz, controlled via the trigger source. According to manufacturer specifications (Exail), the minimum achievable pulse duration of the modulated seed laser is ~30 ps, limited by the bandwidths of both the voltage pulse generator and the intensity modulator.

The optical amplification module consists of three stages pre-amplifier and a power amplifier. All four amplification stages are based on Yb-doped fibers pumped by 976-nm fiber-coupled diode lasers. The 1st pre-amplification stage employs a 1-m-long single-mode Yb-doped fiber with a 6-μm core diameter (PM-YSF-HI-HP, Coherent). The 2nd and 3rd amplification stages use 1-m-long double-clad Yb-doped fibers (PLMA-YDF-10/125, Coherent) to mitigate nonlinear spectral broadening. Between 2nd and 3rd stages, amplified spontaneous emission (ASE) is suppressed using a reflective fiber Bragg grating centered at 1030 nm with a reflection bandwidth of ~0.7 nm (PLMA-YDF-10/125 fiber), connected through an optical circulator. The output of the 3rd stage amplifier is protected with a high-power optical circulator with double cladding pigtail fiber (PLMA-10/125-M, Coherent), to block the back reflection from the fiber output end facet. Laser exiting the 2nd port of the circulator is subsequently mode-matched into a 1 m long PM980 fiber through a mode-field adaptor, which further improves beam quality by suppressing residual higher-order spatial modes of PLMA-10/125-M fiber.

At a repetition rate of 1-3 GHz, the 3rd stage amplifier delivers ~7 W of average power. The beam is collimated using an aspheric lens (C061TMD-B, Thorlabs) and focused through an achromatic lens (AC254-100-B-ML, Thorlabs) into a Yb-doped polarization-maintaining rod-fiber amplifier (aeroGAIN-ROD-MODULE-3.1, NKT Photonics), which has a large mode-field diameter of 65 μm at 1030 nm, 17 dB cladding absorption at 976 nm pump wavelength, and gain centered at 1030 nm. This power amplifier boosts the average output power of 1030 nm laser to a maximum of ~250 W while maintaining near-diffraction-limited beam quality ($M^2$ <1.2) and polarization extinction ratio of ≥15 dB. Between the 3rd stage amplifier and the rod amplifier, a band-pass filter (FLH1030-3, Thorlabs) suppresses residual ASE, a high-power isolator (IO-5-1030-HP, Thorlabs) blocks backward-propagating light, and a half-wave plate aligns the signal polarization with the fast or slow axis of the polarization-maintained rod-fiber amplifier.

The rod-fiber amplifier is counter-pumped by a 530-W diode laser system (DS3-51524-K976BNERN-530.0W, BWT) at 976 nm wavelength. The 976 nm pump laser is delivered through a multimode silica fiber (200-μm core diameter, NA = 0.22). The output is collimated using an aspheric lens (AFL25-25-P-B-285, Asphericon), and focused into the cladding of the Yb-doped rod fiber using a second aspheric lens (AFL25-20-P-B-355, Asphericon). Two dichroic mirrors (108834, Layertec) are used to steer the pump beam and separate the output 1030 nm signal beam from the input 976 nm pump beam. A small fraction (<1%) of the amplified signal beam is extracted using a wedged window (WW41050-B, Thorlabs) for measuring beam profile, spectrum, temporal profile, and monitoring power. Residual 976 nm pump light exiting the 1030 nm signal input end of the rod amplifier is removed using a hole-through parabolic mirror combined with a dichroic mirror. The rod-amplifier and 976 nm pump laser are water cooled.

Fig. S10 shows major characteristics of the 1030 nm pump laser. The laser spectrum in Fig. S10(a) is measured at 250 W signal output power and 3 GHz repetition rate, showing a linewidth of 100 pm. The ratio between the spectrum peak and the background is ~65 dB, indicating the negligible amount of ASE. Fig. S10(b) shows the measured autocorrelation trace with an optical autocorrelator (PulseCheck SM 2000, APE). The duration of the Gaussian fit is measured to be 43 ps at 1 GHz repetition rate, corresponding to 30 ps optical pulse duration. This 30 ps pulse duration is equal to that of seed laser, indicating that the pulse duration experiences negligible compression during the optical amplification due to the large mode areas of Yb-doped fibers being adopted. Fig. S10(c) shows the average power of the 1030 nm laser as a function of the average power of the 976 nm laser diode, where the 1030 nm laser reaches 250 W at 400 W of the diode laser pump. Further power scaling is avoided to prevent potential damage to the rod amplifier.

Fig. S10(d) compares the pulse trains at different repetition rates of 1, 2, and 3 GHz, measured with a photodetector (5 GHz bandwidth, DET08C, Thorlabs). Due to the limited bandwidth, the pulse train at 3 GHz cannot be clearly resolved compared to 1 GHz repetition rate. Fig. S10(e-g) presents the measured radio frequency spectra of the laser pulse train at these three repetition rates. The signal-to-noise ratio decreases from ~64 dB at 1 and 2 GHz to 57 dB at 3 GHz, indicating the compromised stability of the pulse train at 3 GHz repetition rate.

**Section 4. Measurement devices**

The experimental characterization in this work includes measurements of temporal pulse profiles, optical spectra, laser beam profiles, autocorrelation traces, average power, and radio frequency spectrum.

i) Using the experimental configuration shown in Fig. S11(a), the temporal profiles of the pump laser, residual pump, and Raman laser are synchronously monitored with photodetectors to investigate the time-domain behavior of the proposed approach. Temporal profiles at the pump wavelength (~1 μm) are measured using DET08C photodetectors (5 GHz bandwidth, Thorlabs), at 1.9 μm Raman Stokes wavelength are measured with a ET-5000 photodetector (12 GHz bandwidth, Coherent), at the 2.5 μm Raman Stokes wavelength are measured using a PDAVJ8 photodetector (100 MHz bandwidth, Thorlabs). Each photodetector is preceded by a neutral density filter to attenuate the incident light to the linear response range of the detector. Optical filters are also employed to block unwanted wavelengths. All photodetectors are connected to a digital oscilloscope (MSO64B, Tektronix; 6 GHz bandwidth) via SMA/BNC cables rated for 12 GHz bandwidth to suppress signal distortion.

ii) Optical spectra are measured using two spectrometers: Spectro 320 (Instrument Systems) and AQ6317B (ANDO, AssetRelay). Spectra wavelength limited to < 1700 nm are measured using AQ6317B, with a resolution of 10 pm. Spectra involved > 1700 nm are measured with Spectro 320, with a resolution of 0.1 nm at < 632 nm, 0.2 nm in the range of 632 nm to 2000 nm, and 0.4 nm at > 2000 nm. For measurements in the visible and near-infrared regions, laser signals are delivered to the spectrometers via an endless single-mode fiber (ESM-12B, Thorlabs) with custom FC/PC terminations. For mid-infrared measurements, a multimode ZBLAN fiber patch cable (MF11L1AR1, Thorlabs) is used.

iii) Near- and mid-infrared beam profiles are measured with a beam profiler (BP209-IR2, Thorlabs). The beam shape at 721 nm wavelength in Fig. 1(c) is measured with a camera (U3-3680XLE-C-HQ, IDS).

iv) Optical autocorrelation traces are measured using a two-photon absorption-based autocorrelator (PulseCheck SM 2000, APE). The device includes interchangeable detection modules to accommodate different wavelength regions. For the burst mode high repetition rate lasers, due to the low repetition rate of the burst-to-burst signal (1 kHz), the autocorrelator is synchronized with the pump laser to ensure effective signal acquisition. For the continuous mode high repetition rate lasers, we can only roughly measure the autocorrelation traces at $\lesssim$ 1 GHz repetition rate, because the autocorrelator has a < 500 mW limit on the maximum input average power.

v) Average laser power is measured with thermal power meters (S405C, Thorlabs; S425C-L, Thorlabs; UP55G-600F-HD-INT-D0, Gentec-EO).

vi) Radio frequency spectra are measured with a radio frequency analyzer (SA44B, Signal Hound), with 20 kHz resolution.

**Section 5. Design of 1 μm high repetition rate pump laser in burst mode**

Fig. S11(a) shows the experimental configuration for the high repetition rate pump laser in burst mode, comprising a high repetition rate seed laser, a burst modulator and optical amplification module.

The seed laser generates a continuous high-repetition-rate pulse train with the configuration identical to that described in Section 3 of Supplement 1, but the emission wavelength in this part is tunable in the range of 1034-1085 nm. The continuous (sub-)GHz repetition-rate seed laser is subsequently modulated into burst mode using a semiconductor optical amplifier (SOA) modulator (1020SOA-1-5-1, Aerodiode). This SOA functions as an intensity modulator with a rise/fall time of ~2 ns and an extinction ratio of ~40 dB in the 1 μm wavelength range. The resulting optical bursts have a duration of ~20 ns and are shaped into an exponential profile to pre-compensate for gain distortion introduced during optical amplification. The burst-to-burst repetition rate is fixed at 1 kHz.

A linearly polarized Yb-doped fiber amplification module is employed to amplify the burst mode high repetition rate laser. This amplification module consists of three stages of pre-amplifiers, followed by a power amplifier. In the power amplification stage, a polarization-maintaining large-mode-area Yb-doped fiber (PLMA-YDF-25/250-UF, Coherent) is used to mitigate spectral broadening effects. All amplification stages are pumped by 976 nm fiber coupled laser diodes operating in a pulsed regime with pulse durations on the order of hundreds of microseconds, enabling pulsed-pump amplification. This configuration supports broad wavelength tunability from 1034 nm to 1085 nm [3]. High-power optical isolators are used before each stage of amplifier, to block the detrimental back-propagation light. To prevent optical damage at the fiber output due to high burst energy, an endcap is spliced to the fiber tip and coated to reduce back-reflection to <0.1% at both the signal (~1 μm) and pump (976 nm) wavelengths. Due to the absence of bandpass filters within the amplification chain, the output bursts exhibit significant ASE. To suppress ASE while preserving the wide wavelength tunability, two fiber-coupled acousto-optic intensity modulators (200 MHz bandwidth, Csrayzer) are respectively placed before the second and third pre-amplifier stages. The 976 nm pump diodes and acousto-optic modulators are synchronized via a trigger-delay system.

At the output of the pump laser, a bandpass filter (FLH1064-3, Thorlabs) with a 3 nm passband centered at 1064 nm is optionally inserted to suppress ASE. This filter is used to assist the accurate estimation of the average pulse energy via average power measurements. To prevent back-reflection into the power amplification stage, the filter is mounted at a slight angle, which shifts the passband center from 1064 nm to ~1062 nm. Accordingly, the pump laser wavelength is finely tuned to ensure transmission of the pulse train through the filter while maximizing ASE suppression. Representative spectra before and after filtering are shown in Fig. S11(b).

Following the bandpass filter, a C-coated wedged window (WW41050-C, Thorlabs) is inserted to reflect a small fraction (<1%) of the pump laser beam for monitoring the temporal profile with a photodetector (DET08C/M, Thorlabs; 5 GHz bandwidth).

Using the described amplification module, the average pulse energy can be scaled to several hundred nanojoules within a repetition rate range of 500 MHz to 3 GHz, with the measured laser linewidth of approximately 0.13 nm.

Optical autocorrelation measurements show that the amplified optical pulses have an average pulse duration shorter than the specified 30 ps of the seed laser, attributed to self-compression during amplification. The exact pulse duration depends on the power of the 976 nm pump diodes at each amplification stage. To ensure consistency across experiments, the pulse duration is maintained at 23 ps for all repetition rates between 500 MHz and 3 GHz (see Fig. S11(e–f)). This is achieved by appropriately adjusting the pump powers at each (pre-)amplification stage.

A continuous neutral density filter (NDC-50C-4, Thorlabs) is placed after the wedged window for continuous attenuation of the pump laser. Varying the attenuation level does not affect the pulse duration of 23 ps, as the peak intensity of the collimated beam remains low (~0.1 MW/cm$^2$). Note that the 23 ps pulse duration is measured after this neutral density filter.

The maximum average power measured after the neutral density filter is 10.8 mW at 500 MHz, 17.3 mW at 1 GHz, 27.8 mW at 2 GHz, and 34.0 mW at 3 GHz repetition rate. By counting pulses exceeding 10% of the average background voltage, the number of pulses per ~19 ns burst is estimated to be approximately 11, 22, 41, and 71 at repetition rates of 500 MHz, 1 GHz, 2 GHz, and 3 GHz, respectively. The calculated average pulse energies are 982, 786, 678, and 478 nJ, respectively. The pump wavelength is tunable from 1034 to 1085 nm, but is primarily fixed at 1062 nm for producing experimental data in Fig. 3.

To ensure thermal stability of the amplification module (see Fig. S11(g)), fan cooling is implemented.

**Section 6: One-dimensional coupled Maxwell-Bloch equations**

In this work, we adopt the fiber based one-dimensional unidirectional Maxwell-Bloch equations provided in [4]. For simplicity, only the pump and the first-order Raman Stokes are considered. Under this approximation, the Maxwell-Bloch equations (S1) and (S2) in Ref. [4] are expressed as

$$\frac{\partial E_s}{\partial z\prime} + \frac{1}{v_s}\frac{\partial E_s}{\partial t} = -ik_2 Q^* E_p \exp(i(\beta_p - \beta_s)z) - \frac{1}{2}\alpha_s E_s$$
$$\frac{\partial E_p}{\partial z} + \frac{1}{v_p}\frac{\partial E_p}{\partial t\prime} = -i\frac{\lambda_s}{\lambda_p} k_2 Q E_s \exp(i(\beta_s - \beta_p)z) - \frac{1}{2}\alpha_p E_p \tag{1}$$
$$\frac{\partial Q}{\partial t'} + \frac{Q}{T_2} = -\frac{1}{4} ik_1 E_p E_s^* \exp(i(\beta_p - \beta_s)z)$$

where $E_p$ and $E_s$ are slowly varying envelopes of the pump and Stokes waves in the unit of V/m. They are coupled through molecular coherence $Q$ (dimensionless unit). $z$ is the propagation distance along fiber, and $t$ is time. Compared with Ref. [4], the time-dependent terms of $\frac{\partial E_s}{\partial t}$ and $\frac{\partial E_p}{\partial t}$ are added, since the pump/Raman waves in this work are in the form of (ultrafast) pulses. $v_s$ and $v_p$ stands for the group velocities of the pump and Stokes lasers [5]. $\beta_p$ and $\beta_s$ are the propagation constants of the pump and Stokes lasers. $k_1$ and $k_2$ are coupling constants. $\alpha_p$ and $\alpha_s$ are optical losses ($m^{-1}$) of the pump and Raman waves. $\lambda_p$ and $\lambda_s$ are wavelengths of the pump and Stokes waves. $T_2$ is the dephasing time of molecular vibration, which links to the spectral width Δν (FWHM) of the Raman gain through $T_2 = 1/(\pi\Delta\nu)$. The coupling constants $k_1$ and $k_2$ can be expressed as

$$k_1 = \sqrt{\frac{2\gamma_g c^2 \varepsilon_0^2}{N T_2 \hbar\omega_s}} \tag{2}$$

$$k_2 = N\hbar\omega_s k_1 / c^2 \frac{N\hbar\omega_s k_1}{2\varepsilon_0 c} \tag{3}$$

where $\gamma_g$ is the Raman gain coefficient of the Q(1) vibrational mode of $H_2$, which depends on pump wavelength, $H_2$ pressure, and temperature [6]. $N$ is the gas molecular number density, equal to $P/k_B T$ based on ideal gas law, where P is pressure with the unit of Pa, $k_B$ is Boltzmann constant, T is absolute temperature (K). $\hbar$ is Planck's constant, $\omega_s$ and $v_s$ are the angular frequency and group velocity of the Stokes wave, respectively. c is light speed.

To facilitate the simulation, the equations are transformed to a moving frame using the Raman group velocity $v_s$ as a reference, i. e., $t' = t - z/v_s$, and $z' = z - v_s t$. In this case, the equation (1) can be expressed as

$$\frac{\partial E_s}{\partial z\prime} = -ik_2 Q^* E_p \exp(i(\beta_p - \beta_s)z') - \frac{1}{2}\alpha_s E_s$$
$$\frac{\partial E_p}{\partial z\prime} + (\frac{1}{v_p} - \frac{1}{v_s})\frac{\partial E_p}{\partial t} = -i\frac{\lambda_s}{\lambda_p} k_2 Q E_s \exp(i(\beta_s - \beta_p)z') - \frac{1}{2}\alpha_p E_p \tag{4}$$
$$\frac{\partial Q}{\partial t'} + \frac{Q}{T_2} = -ik_1 E_p E_s^* \exp(i(\beta_p - \beta_s)z')$$

According to the property of $H_2$-filled ARHCF-2 provided in Section 2 of Supplement 1, $\alpha_p$ and $\alpha_s$ are set as 0.01266 $m^{-1}$ (55 dB/km) and 0.002763 $m^{-1}$ (12 dB/km). The group velocities are calculated to be $2.99165976712772 \times 10^8$ m/s and $2.99028797605555 \times 10^8$ m/s at 16 bar $H_2$-filled ARHCF-2. Pump and Raman wavelengths are set as 1062 nm and 1.9 µm, respectively. The value of Δν of the Q(1) vibrational Raman of $H_2$ can be calculated from Ref. [6], to calculate the dephasing time $T_2$ (see Fig. S1). Temperature is set to 298 K throughout the simulation. A uniform electric field floor of 500 V/m is set as the seed of Raman signal, this is approximately in the quantum noise level [4]. The initial pump burst $E_p(t', z')$ is expressed as

$$E_p(t', z') = \sqrt{\frac{2}{c\varepsilon_0}\frac{P_0}{A_{eff}}} \exp(-2\ln(2)\frac{\left(t' - \frac{1}{frep}\right)(n - floo\ \ (\frac{1}{2}N_{pulse}))}{T_{FWHM}}) \tag{5}$$

where $A_{eff}$ is the mode field diameter of the ARHCF-2, which is 27 µm according to the simulation in Section 2 of Supplement 1. $P_0$, $f_{rep}$, n, $N_{pulse}$, and $T_{FWHM}$ stand for the peak power, repetition rate, $n^{th}$ pulse, total pulse number, and FWHM pulse duration of the pump.

To solve equation (4), $E_p$ , $E_s$ and $Q$ are respectively expressed in the form of $A_p(t', z')e^{i\varphi_p(t', z')}$ , $A_s(t', z')e^{i\varphi_s(t', z')}$, $A_q(t', z')e^{i\varphi_q(t', z')}$. Then, the real and imaginary parts of equation (4) are separated as

$$
\begin{aligned}
\frac{\partial A_s}{\partial z'} &= k_2 A_q A_p \sin\left(\varphi_p - \varphi_q - \varphi_s + (\beta_p - \beta_s)z\right) - \frac{1}{2}\alpha_s A_s \\
A_s \frac{\partial \varphi_s}{\partial z'} &= -k_2 A_q A_p \cos(\varphi_p - \varphi_q - \varphi_s + (\beta_p - \beta_s)z) \\
\frac{\partial A_p}{\partial z'} + \left(\frac{1}{v_p} - \frac{1}{v_s}\right)\frac{\partial A_p}{\partial t'} &= \frac{\lambda_s}{\lambda_p} k_2 A_q A_s \sin\left(-(\varphi_p - \varphi_q - \varphi_s) - (\beta_p - \beta_s)z\right) - \frac{1}{2}\alpha_p A_p \\
A_p \frac{\partial \varphi_p}{\partial z'} + \left(\frac{1}{v_p} - \frac{1}{v_s}\right) A_p \frac{\partial \varphi_p}{\partial t'} &= -\frac{\lambda_s}{\lambda_p} k_2 A_q A_s \cos\left(-(\varphi_p - \varphi_q - \varphi_s) - (\beta_p - \beta_s)z\right)
\end{aligned} \tag{6}
$$

$Q$ is expressed as:

$$
Q(t', z') = -ik_1 \int_{T_{min}}^{t'} E_p(t'', z') E_s^*(t'', z') \exp(i(\beta_p - \beta_s)z') e^{-\frac{1}{T_2}(t'-t'')} dt'' \tag{7}
$$

where $T_{min}$ is the minimum value of the time window in the simulation. In the numerical solution, the 'method of lines' is used to transform Equation (6) to ordinary differential equations, which are then solved with midpoint Euler method [7].

The simulation results presented in Fig. 2 are generated using a temporal step size of 488 fs and a total time window of 32 ns.

Note that other nonlinear effects (e.g., self-phase modulation and gas ionization) are neglected in simulations. This is because the low pump pulse peak power of only ~4 kW is orders of magnitude below the threshold of other nonlinear effects. Dispersion-caused pulse broadening/compression effects are also neglected, due to the long pulse duration of 23 ps and narrow laser linewidth of 0.13 nm.

**Section 7: Supplementary details to Fig. 3**

Here we provide supplementary details about the experimental validation of MACMO in burst mode.

In this experiment, the pump laser (see Section 5 of Supplement 1) is coupled into $H_2$-filled ARHCF-2, which is sealed with gas cells at both ends. The pump polarization is optimized using a half-wave plate to maximize the vibrational Raman conversion efficiency. The coupling efficiency into the ARHCF-2 is approximately 80%. The output beam from the ARHCF-2 is collimated, and a dichroic mirror is used to separate the generated Raman signal from the residual pump. Photodetectors synchronously record the temporal profiles of the input pump pulses, the residual pump, and the Raman pulses.

The traces in Fig. 3(a) have been temporally aligned to compensate for unequal signal propagation delays in the detection system (see Visualization 1).

Fig. S7(b) shows the evolution of quantum efficiency as a function of average pump pulse energy at repetition rates from 500 MHz to 3 GHz, with the $H_2$ pressure fixed at 16 bar. At a repetition rate of 3 GHz, the threshold for generating 1.9 μm Raman pulses is reached at an average pump pulse energy as low as ~120 nJ. The maximum average quantum efficiency of 23% is obtained at a pump energy of only ~200 nJ, corresponding to an average power of ~600 W for continuous operation at the same repetition rate. As the repetition rate decreases, the SRS threshold increases and progressively higher pump energies are required, reaching ~900 nJ at 500 MHz.

Fig. S7(c) shows the average quantum efficiency of the 1.9 μm Raman signal as a function of $H_2$ pressure for different pump repetition rates, with the pump pulse energy set to its maximum value at each repetition rate. As the $H_2$ pressure increases, the Raman gain coefficient rises and transient Raman scattering is increasingly suppressed, leading to a corresponding growth in quantum efficiency. Beyond certain pressures, however, the quantum efficiency reaches a maximum and then gradually decreases. This decline can arise from three factors: (i) At higher pressures, the increased Raman gain coefficient shifts the point of maximum conversion to shorter fiber lengths, leaving a longer residual fiber section where fiber propagation losses dominate; (ii) The contribution of the MACMO effect may diminish due to faster molecular dephasing; (iii) Partial back-conversion of the 1.9 μm Raman signal to the 1062 nm pump wavelength may occur via the 1st order Raman anti-Stokes generation. Consistent with the latter two effects, the residual pump power increases with $H_2$ pressure. For example, at a repetition rate of 3 GHz, the residual pump power increases monotonically from 2.79 mW at 15 bar to 3.14 mW at 30 bar (see Fig. S7(d)).

At a repetition rate of 500 MHz, a quantum efficiency of ~5% is still achieved even though the 2 ns pulse-to-pulse interval is nearly an order of magnitude longer than the ~0.2 ns dephasing time of $H_2$ at pressures above 10 bar. This behavior can be attributed to three favorable factors associated with high $H_2$ pressure: an enhanced Raman gain coefficient, suppression of the transient Raman scattering regime, and a high average pump pulse energy of 982 nJ. Under these conditions, conventional SRS may also contribute to the observed conversion because pulse energies approach the microjoule regime. Nevertheless, when the pump pulse energy is reduced to 800 nJ, the Raman signal vanishes, indicating that MACMO at least contributes at repetition rates of ≥1 GHz, where the average pump pulse energy remains below 800 nJ.

It is worth noting that, although both the simulations (Fig. 2) and experiments (Fig. 3) support the validity of the MACMO, the detailed results exhibit quantitative discrepancies. For example, at 16 bar $H_2$ pressure and an average pump pulse energy of 100 nJ at a 3 GHz repetition rate, simulations yield an average quantum efficiency of 62%. However, experimentally, the SRS threshold is approximately 120 nJ (Fig. S7(b)). This deviation may arise from several contributing factors, detailed below:

i) The numerical model presented in Section 6 of Supplement 1 does not account for the influence of pump laser linewidth on SRS efficiency. When the pump linewidth substantially exceeds the width of the Raman gain profile, the conversion efficiency can be reduced, particularly in the presence of dispersion-induced walk-off between the pump and Raman pulses [8]. In our experiment, the pump linewidth is 130 pm, whereas the calculated width of the Q(1) Raman gain profile in $H_2$ is only 848 MHz (equivalent to 3 pm at 1062 nm) at 16 bar pressure, indicating a significant mismatch.

ii) The simulation assumes uniform pump pulses within each burst (see top panel of Fig. 2(d)). In contrast, pulse energies vary across the burst in experiments (see Fig. 3(a)).

iii) The simulated loss values of the ARHCF-2 may be inaccurate due to variations in capillary wall thickness along the fiber length. Direct measurement of fiber loss via the cut-back method is not performed, as the available ARHCF-2 sample is of limited length and reserved for other experiments.

iv) The Yb-doped fiber used in the power amplification stage (PLMA-YDF-25/250-UF, Coherent) supports multiple transverse modes at ~1 µm. Although higher-order modes are largely suppressed by coiling the fiber to a ~10 cm diameter, partial excitation may still occur. The presence of these modes can degrade Raman quantum efficiency.

v) The simulation does not account for fiber coupling losses from the pump to the ARHCF. Experimentally, this loss is approximately 20%.

**Section 8: Exploitation of MACMO in the mid-infrared and visible regions**

To assess the spectral scalability of MACMO, we investigate frequency conversion toward both the visible and mid-infrared spectral regions. In the first experiment, a burst-mode pump laser operating at a repetition rate of 3 GHz with an average pulse energy of 478 nJ is used to pump $H_2$-filled ARHCF-2 at a pressure of 30 bar. This $H_2$ pressure enables the observation of 1$^{st}$ and 2$^{nd}$ order Q(1) Raman anti-Stokes lines at 737 nm and 534 nm, respectively. The measured output spectrum is shown in Fig. S12(a). Although the average power of the generated anti-Stokes signals is very low (estimated < 10 µW) due to residual phase mismatch, this result demonstrates the feasibility of extending MACMO-driven frequency conversion to the visible and potentially even the ultraviolet spectral regions.

In the second experiment, we demonstrate wavelength tunable high-repetition-rate mid-infrared Raman generation at a repetition rate of 3 GHz. The experimental configuration consists of two cascaded ARHCFs. In the 1$^{st}$ stage, a $CO_2$-filled ARHCF-3 is used to convert the 1 µm pump into the 1.2 µm spectral region through the 1$^{st}$ order $\nu_1/2\nu_2$ vibrational Raman Stokes generation in $CO_2$. The resulting 1.2 µm Raman output then serves as the pump for the 2$^{nd}$ stage, which employs an $H_2$-filled ARHCF-4 to generate 1$^{st}$ order Q(1) vibrational Raman Stokes in the 2.5 µm mid-infrared region. Figure S12(b) shows that tuning the pump wavelength from 1034 to 1085 nm produces tunable Raman lines between approximately 1207 and 1277 nm in the 1$^{st}$ stage ARHCF. The corresponding average pulse energy is ~200 nJ, with approximately 54 pulses per burst. When this output is coupled into the 2$^{nd}$ stage $H_2$-filled ARHCF-4, tunable mid-infrared Raman pulse generation at a repetition rate of 3 GHz is obtained in the 2.42–2.54 µm wavelength range, corresponding to pump wavelengths between 1034 nm and 1055 nm (Fig. S12(c)). No mid-infrared Raman output is observed for pump wavelengths above ~1055 nm, which is attributed to increasing propagation loss in the 2$^{nd}$ stage ARHCF-4 (see Fig. S12(d-f)).

In addition to the vibrational Raman lines, Fig. S12(c) also shows clear S(1) rotational Raman (anti-)Stokes generation of $H_2$, which is absent in the 1.9 µm Raman laser experiments (see Fig. S12(a)). This behavior is attributed to the compromised linear polarization extinction ratio of approximately 11 dB in the 1.2 µm Raman pump, which facilitates the excitation of rotational Raman transitions.

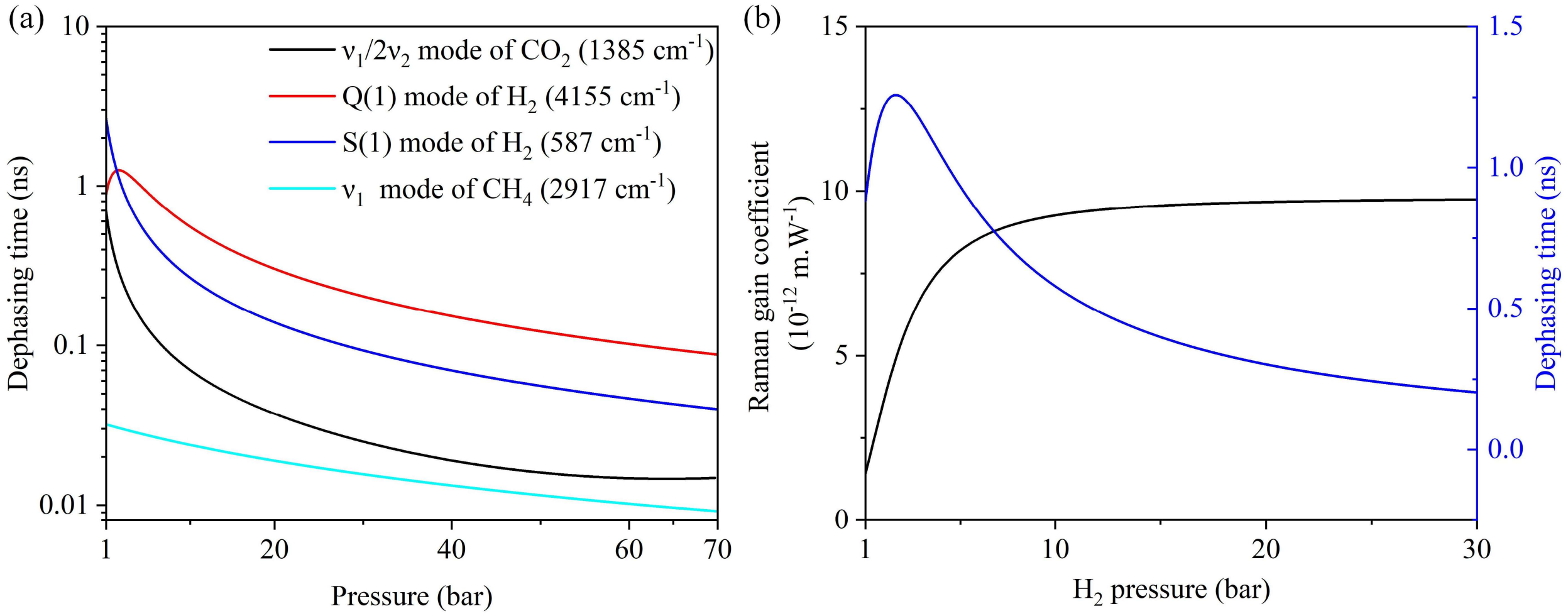

Fig. S1. Dephasing time and Raman gain of gas media. (a) Molecular dephasing times for representative vibrational and rotational modes of $CO_2$, $H_2$, and $CH_4$ as a function of pressure. The dephasing time is calculated as $1/(\pi\Delta\upsilon)$, where $\Delta\upsilon$ is the Raman gain linewidth of the selected mode. Linewidth values are extracted from Ref. [9–11]. Raman shift coefficients (in $cm^{-1}$) are indicated in parentheses in the legend for each mode. (b) Raman gain coefficient and corresponding dephasing time for the Q(1) vibrational mode of $H_2$. The dephasing time is identical to that shown in (a). The Raman gain coefficient is calculated following Ref. [6] for a pump wavelength of 1062 nm.

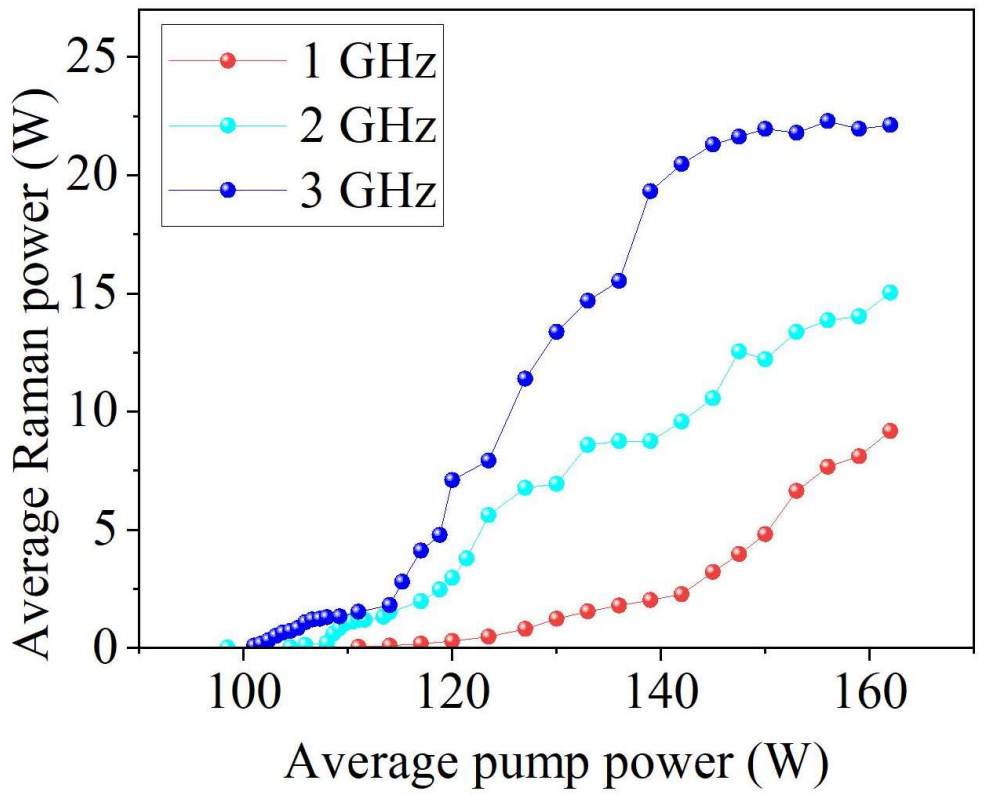

Fig. S2. Average power of the 1.8 μm vibrational Raman Stokes emission as a function of average pump power at different repetition rates. This figure complements Fig. 1(b).

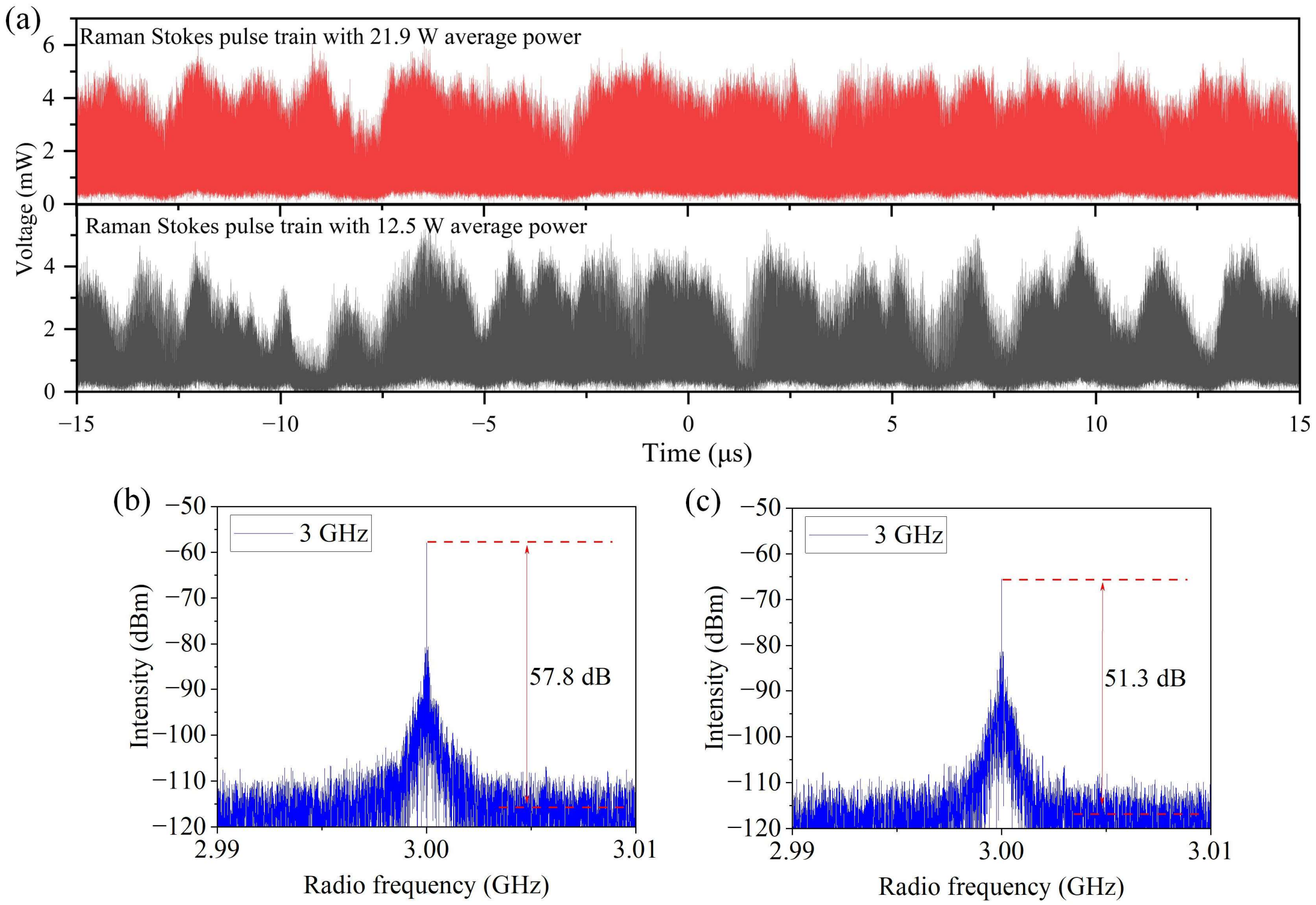

Fig. S3. Power-dependent stability of the 1.8 μm Raman pulse train. (a) Raman pulse trains at average powers of 21.9 W (top) and 12.5 W (bottom). (b, c) Corresponding radio-frequency spectra at 21.9 W (b) and 12.5 W (c). Data in top of (a) corresponds to Fig. 1(e), and data in (b) corresponds to Fig. 1(g).

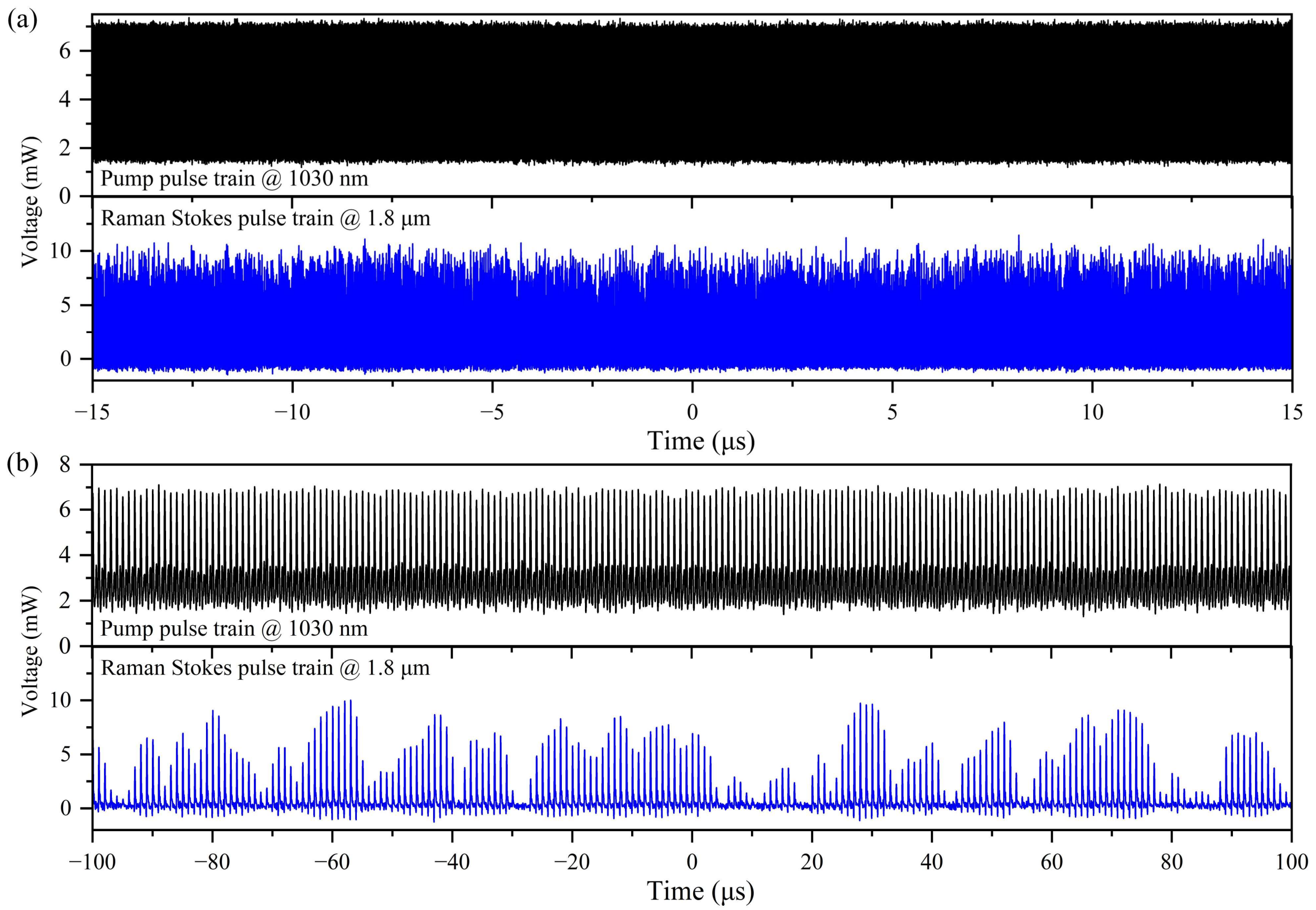

Fig. S4. Temporal characteristics at 1 GHz repetition rate. (a) Raman Stokes (bottom) at 9.2 nJ pulse energy and corresponding pump (top) at 162 nJ, measured over a 30 µs window. (b) Temporal zoom of the pulse trains in (a).

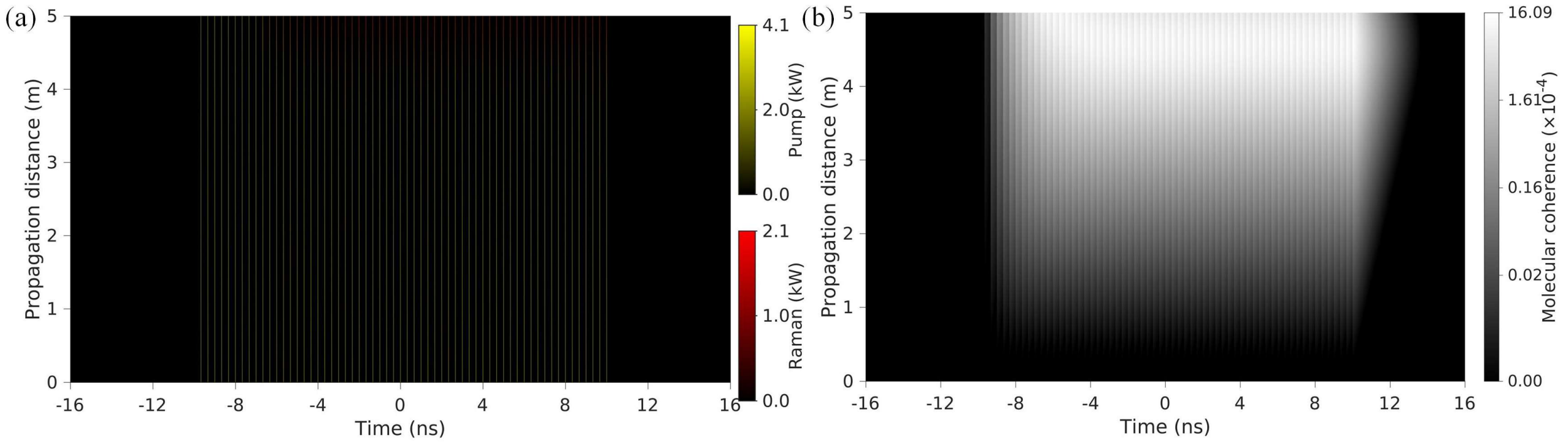

Fig. S5. Simulated temporal evolution of the pump, Raman, and molecular coherence along the 16 bar $H_2$-filled ARHCF-2. This figure complements Fig. 2(a) and 2(c) by showing the full evolution of all pump and Raman pulses. (a) Power evolution of pump pulses (yellow) and frequency-converted Raman pulses (red) along the fiber. (b) Logarithmic-scale amplitude evolution of molecular coherence.

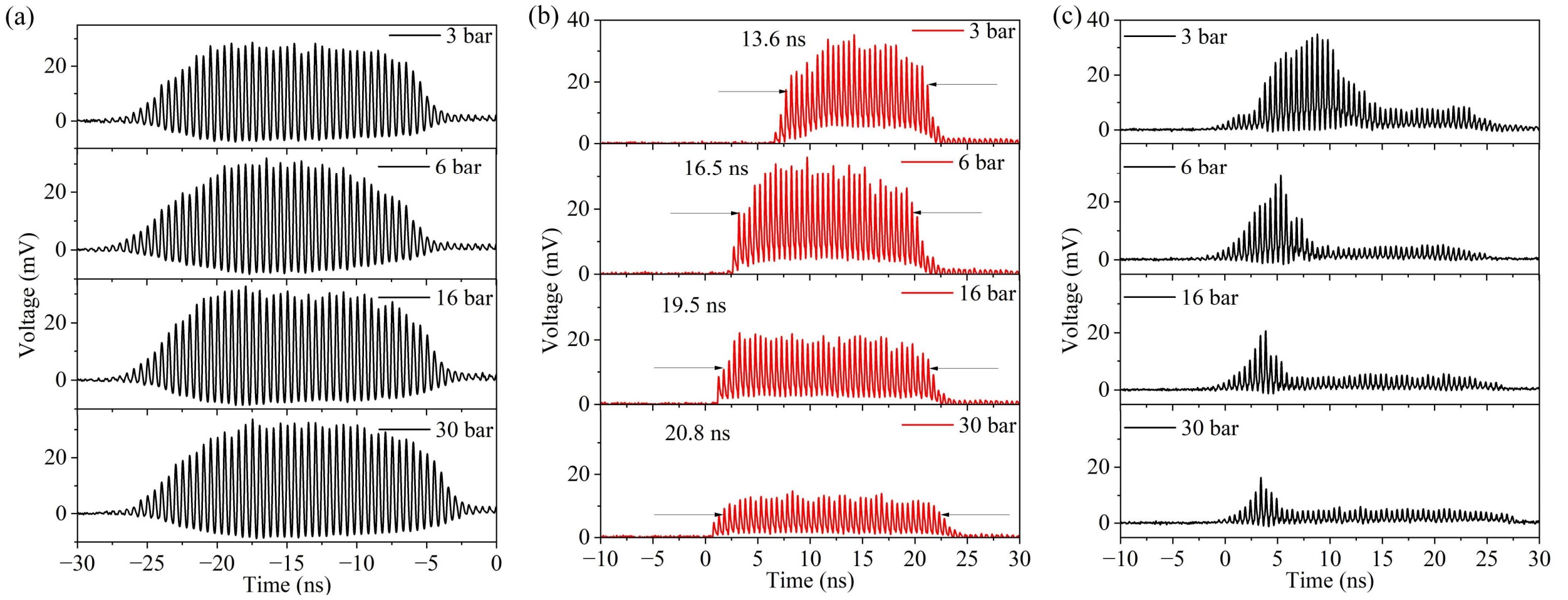

Fig. S6. Temporal profiles of pump, Raman, and residual pump signals at different $H_2$ pressures, measured at a repetition rate of 3 GHz. (a) Pump pulse trains. (b) Corresponding Raman signals. (c) Residual pump signals corresponding to (a).

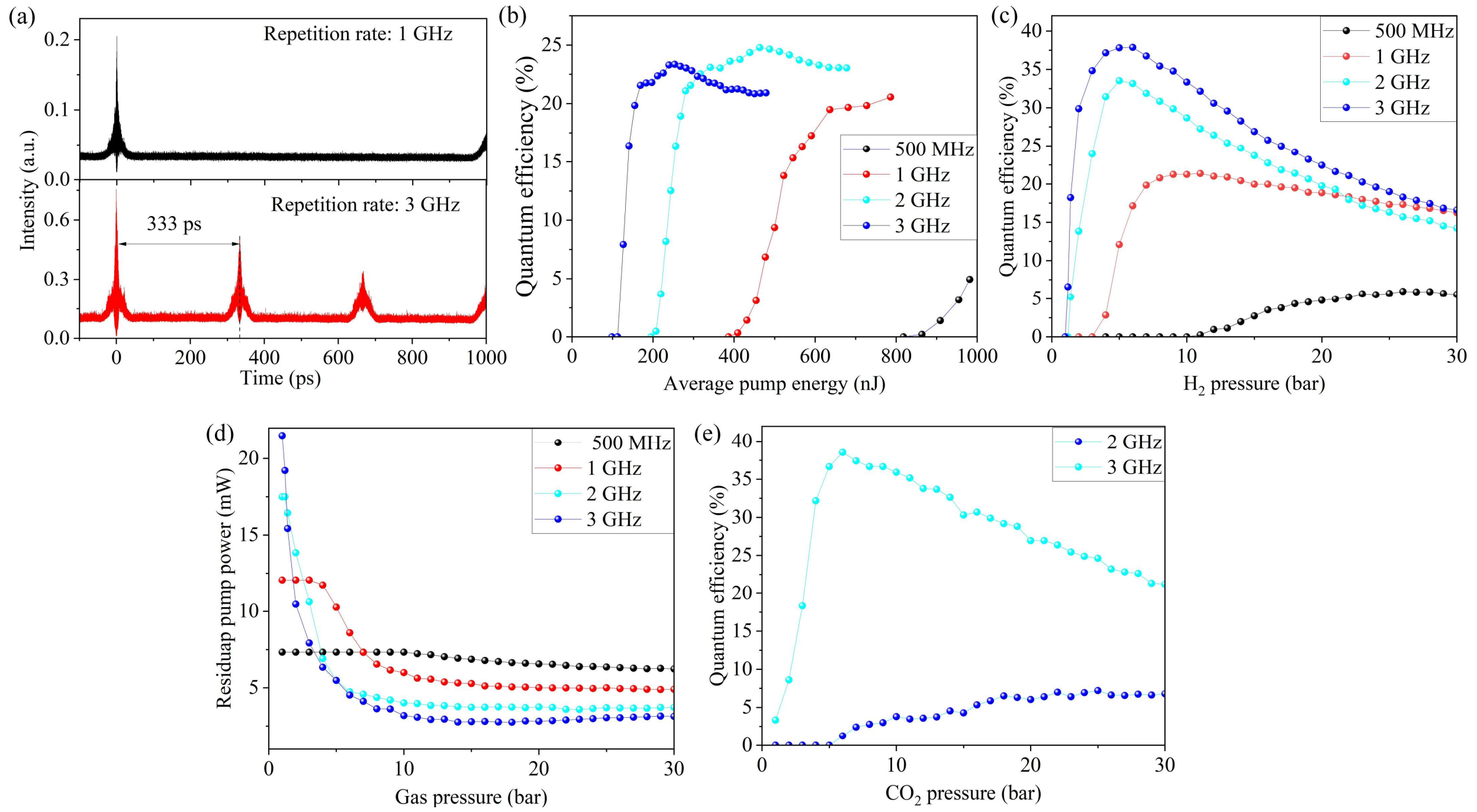

Fig. S7. Supplementary characterization of burst-mode Raman frequency conversion. (a) Autocorrelation traces of the 1.9 μm Raman emission over a 1.1 ns window. This figure corresponds to Fig. 3(b). (b) Quantum efficiency versus average pump pulse energy at repetition rates from 500 MHz to 3 GHz in 16 bar $H_2$-filled ARHCF-2. (c) Quantum efficiencies of the 1.9 μm Raman burst versus $H_2$ pressure, at repetition rates from 500 MHz to 3 GHz. (d) Corresponding residual pump power versus pressure. (e) Quantum efficiency of 1.25 μm Raman emission in $CO_2$-filled ARHCF-2, at the repetition rate of 2 GHz, and 3 GHz, respectively. $CO_2$ pressure is fixed at 2 bar.

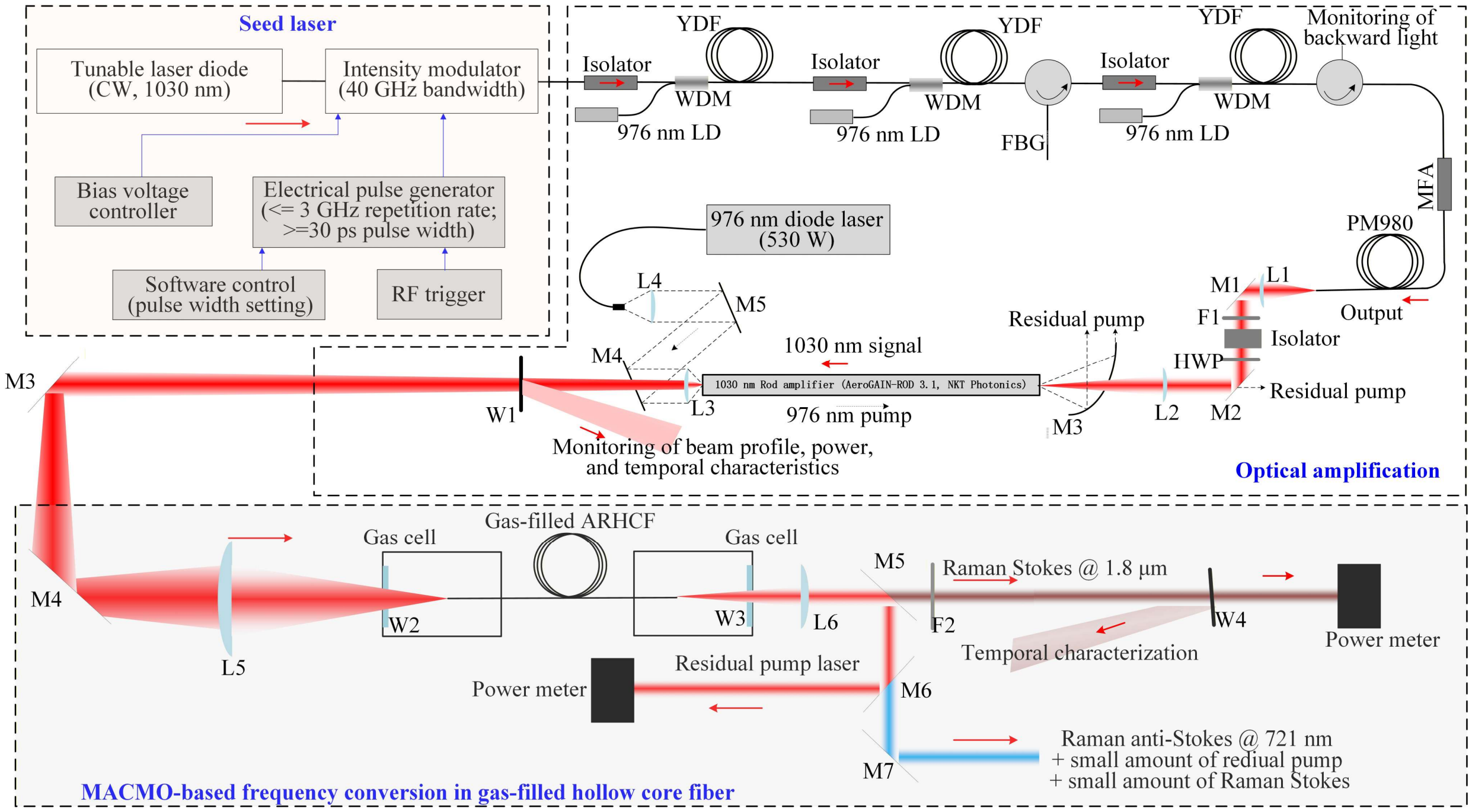

Fig. S8. Experimental setup for continuous GHz MACMO-based frequency conversion. Schematic of the system converting a 1030 nm pump to 1.8 µm Raman emission. The setup includes a seed laser, high-power amplifier, and $H_2$-filled ARHCF-1. M1–M7: mirrors (M2, M5–M7 dichroic); L1–L6: lenses; W1–W4: optical windows; F1: 1030 nm bandpass filter; F2: 1.3 µm long-pass filter; LD: laser diode. Water cooling is omitted for clarity. See Sections 1 and 3 of Supplement 1 for details.

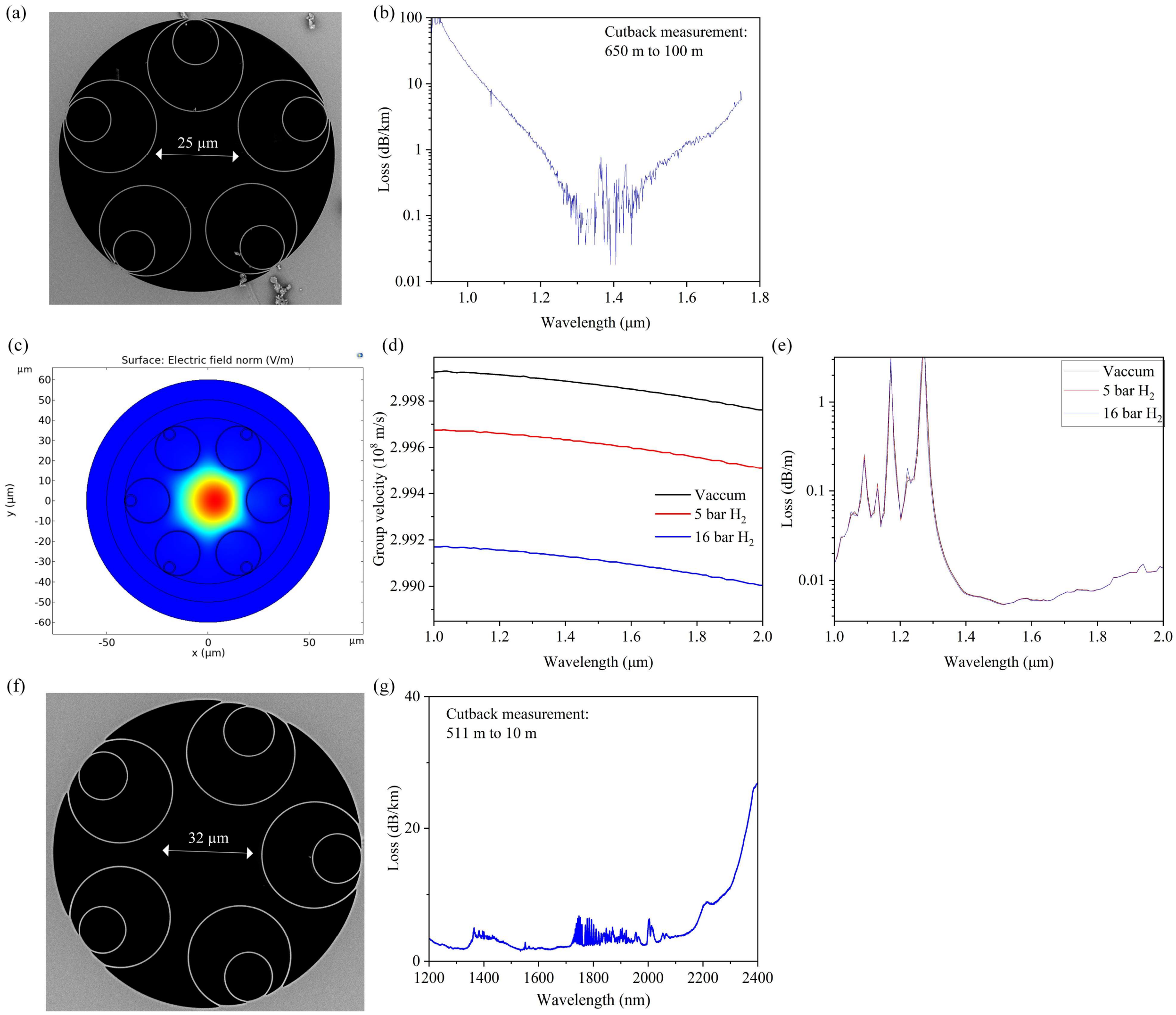

Fig. S9. | Characterization of ARHCF-1, ARHCF-2, and ARHCF-4. (a, b) Experimental characterization of ARHCF-1: scanning electron microscope image (a) and transmission loss (b). (c-e) Simulated properties of ARHCF-2: mode field distribution at 1062 nm (c), group velocity (d), and loss spectra (e). (f, g) Experimental characterization of ARHCF-4: scanning electron microscope image (f) and transmission loss (g).

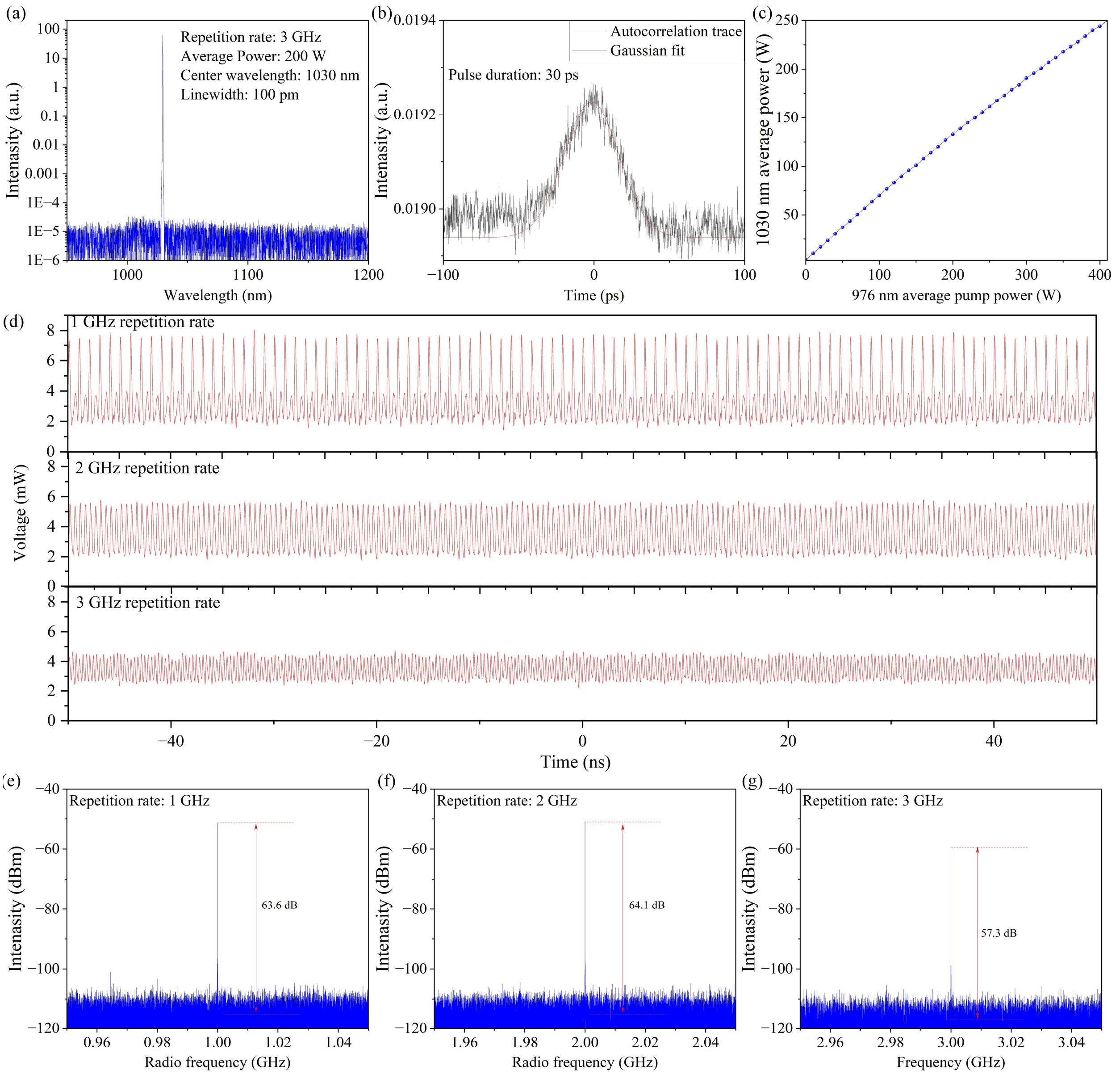

Fig. S10. Characteristics of the 1030 nm pump laser. (a) Optical spectrum at 3 GHz (10 pm resolution). (b) Autocorrelation trace at 1 GHz. (c) Average power of 1030 nm laser output from the Yb-doped rod amplifer, versus 976 nm pump diode power. (d) Pulse trains at different repetition rates at 160 W average power. (e, f) Corresponding radio-frequency spectra.

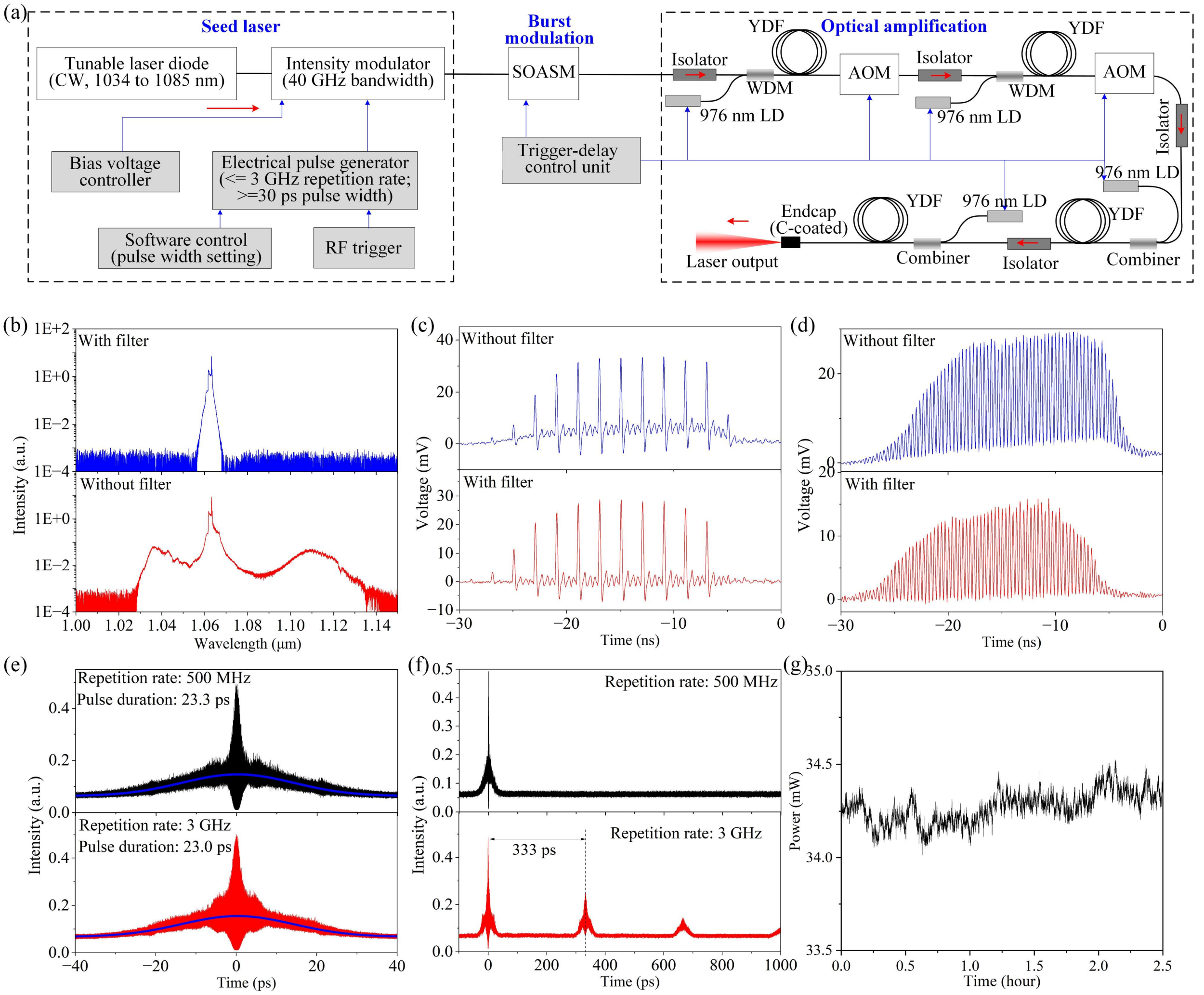

Fig. S11. Configuration and characterization of burst-mode pump source. (a) Experimental setup. (b) Output spectra at 1 GHz with and without a 1062 nm bandpass filter. (c, d) Temporal profiles at 500 MHz (c) and 3 GHz (d). (e, f) Autocorrelation traces with 80 ps (e) and 1.1 ns (f) scanning ranges. (g) Stability monitoring of the pump average power over 2.5 hours at 3 GHz repetition rate, with the 1062 nm bandpass filter applied.

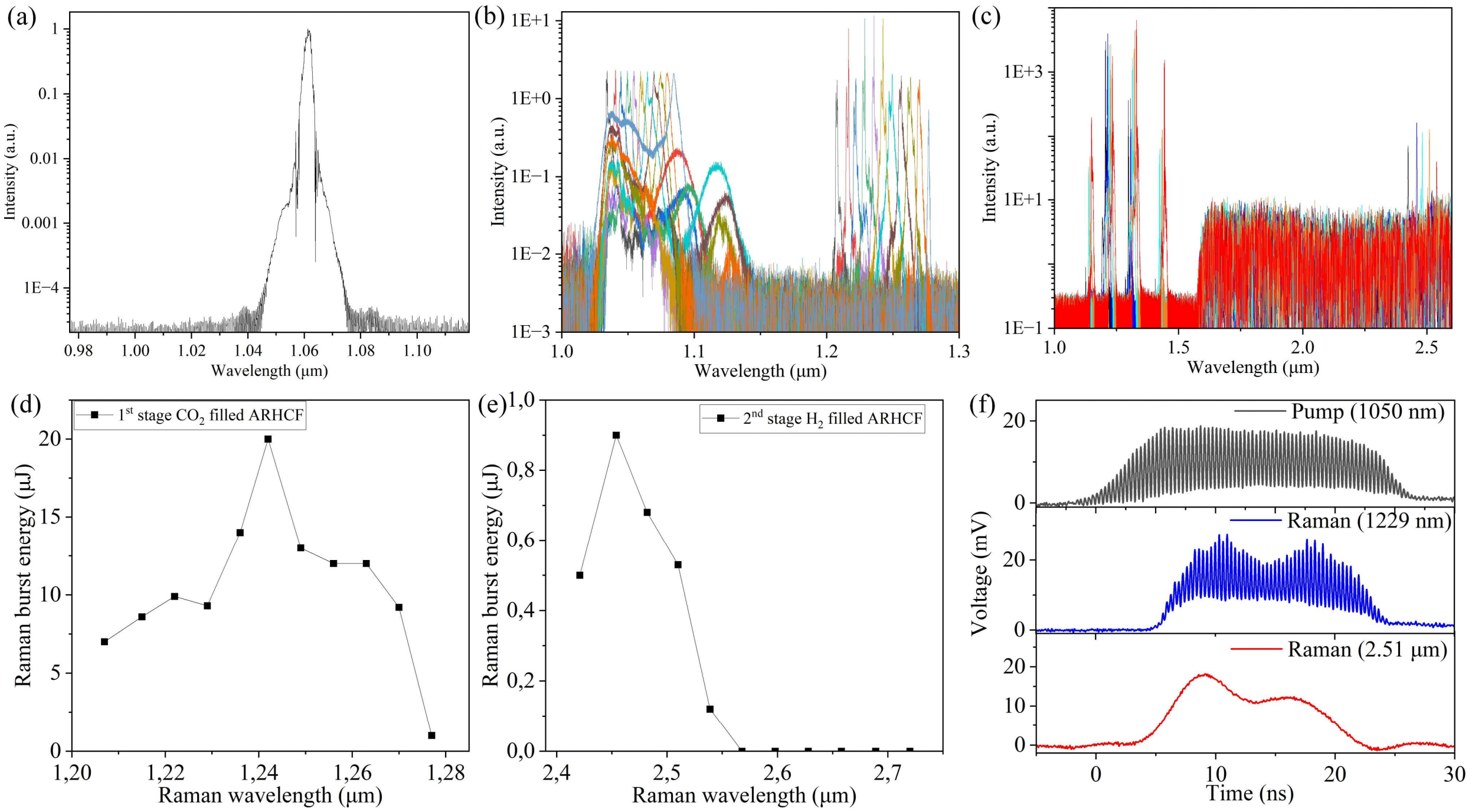

Fig. S12. Broadband Raman frequency conversion enabled by MACMO. (a) Visible anti-Stokes emission via the Q(1) vibrational mode in $H_2$-filled ARHCF-2 at 30 bar pressure. (b, c) Mid-infrared wavelength-tunable Raman Stokes generation using a cascaded configuration: $CO_2$-filled ARHCF-3 (1st stage) followed by $H_2$-filled ARHCF-4 (2nd stage). (b) Output spectra measured from the 1st stage $CO_2$-filled ARHCF-3, obtained by tuning the pump wavelength from 1034 to 1085 nm. (c) Output spectra from 2nd stage $H_2$-filled ARHCF-4. (d) Measured burst energy near 1.2 μm from the 1st stage $CO_2$-filled ARHCF-3. (e) Measured burst energy near ~2.5 μm region from the 2nd stage $H_2$-filled ARHCF-4. (f) Temporal profiles of the pump burst (1050 nm), 1229 nm Raman burst (stage 1), and 2.51 μm Raman burst (stage 2). Individual pulses at 2.5 μm are unresolved due to detector bandwidth limitations (100 MHz).

## Reference list

1. Y. Wang, L. Hong, C. Zhang, J. Wahlen, J. E. Antonio-Lopez, M. K. Dasa, A. I. Adamu, R. Amezcua-Correa, and C. Markos, "Synthesizing gas-filled anti-resonant hollow-core fiber Raman lines enables access to the molecular fingerprint region," Nat. Commun. 15, 9427 (2024).
2. A. I. Adamu, Y. Wang, Md. S. Habib, Manoj. K. Dasa, J. E. Antonio-Lopez, R. Amezcua-Correa, O. Bang, and C. Markos, "Multi-wavelength high-energy gas-filled fiber Raman laser spanning from 1.53 µm to 2.4 µm," Opt. Lett. 46, 452–455 (2021).
3. Y. Wang, et al. "Tunable, high pulse energy and narrow linewidth gas-filled fiber laser across near- and mid-infrared," Laser & Photonics Rev. e71703 (2026).
4. S. T. Bauerschmidt, D. Novoa, A. Abdolvand, and P. St. J. Russell, "Broadband-tunable LP01 mode frequency shifting by Raman coherence waves in a $H_2$-filled hollow-core photonic crystal fiber," Optica 2, 536–539 (2015).
5. M. G. Raymer and I. A. Walmsley, "III The quantum coherence properties of stimulated Raman scattering," in Progress in Optics, E. Wolf, ed. (Elsevier, 1990), Vol. 28, pp. 181–270.
6. W. K. Bischel and M. J. Dyer, "Wavelength dependence of the absolute Raman gain coefficient for the Q(1) transition in $H_2$," J. Opt. Soc. Am. B 3, 677–682 (1986).
7. G. Hilfer and C. R. Menyuk, "Stimulated Raman scattering in the transient limit," J. Opt. Soc. Am. B 7, 739–749 (1990).
8. W. Trutna, Y. Park, and R. Byer, "The dependence of Raman gain on pump laser bandwidth," IEEE J. Quantum Electron. 15, 648–655 (1979).
9. D. Hanna, D. Pointer, and D. Pratt, "Stimulated Raman scattering of picosecond light pulses in hydrogen, deuterium, and methane," IEEE J. Quantum Electron. 22, 332–336 (1986).
10. G. C. Herring, M. J. Dyer, and W. K. Bischel, "Temperature and density dependence of the linewidths and line shifts of the rotational Raman lines in $N_2$ and $H_2$," Phys. Rev. A (Coll. Park). 34, 1944–1951 (1986).
11. J. Baran, A. Grofcsik, and W. J. Jones, " Motional narrowing in the $\nu_1/2\nu_2$ Fermi resonance diad of $CO_2$," Mol. Phys. 45, 1291–1297 (1982).